\documentclass[12pt]{iopart}

\usepackage{iopams}

\usepackage{graphicx}
\usepackage{epstopdf}

\usepackage[mediumspace,mediumqspace,squaren]{SIunits}

\usepackage{subcaption}
\usepackage{subfig}

\begin{document}

\title[Fast-ion redistribution in MAST]{Measurements and modelling of fast-ion redistribution due to MHD instabilities in the Mega-Amp Spherical Tokamak}
\author{O M Jones$^{1,2}$, C A Michael$^2$\footnote{Present address: Plasma Research Laboratory, Research School of Physical Science and Engineering, Australian National University, Canberra, ACT 0200, Australia}, M Cecconello$^3$, I Wodniak$^3$, K G McClements$^2$, D L Keeling$^2$, C D Challis$^2$, M Turnyanskiy$^4$, A J Meakins$^2$, N J Conway$^2$, B J Crowley$^2$\footnote{Present address: General Atomics, San Diego, CA 92121-1122, USA}, R J Akers$^2$ and the MAST team$^2$}
\address{$^1$Department of Physics, Durham University, South Road, Durham DH1 3LE, UK}
\address{$^2$EURATOM/CCFE Fusion Association, Culham Science Centre, Abingdon, Oxon. OX14 3DB, UK}
\address{$^3$Department of Physics and Astronomy, Uppsala University, SE-751 05 Uppsala, Sweden}
\address{$^4$ITER Physics Department, EFDA CSU Garching, Boltzmannstra{\ss}e 2, D-85748 Garching, Germany}

\ead{owen.jones@ccfe.ac.uk}

\begin{abstract}
The results of a comprehensive investigation into the effects of various MHD modes on the NBI-generated fast-ion population in MAST plasmas are reported. Fast-ion redistribution due to low-frequency ($20-50\usk\kilo\hertz$) chirping energetic particle modes known as fishbones, as well as the long-lived internal kink mode, is observed with the Fast-Ion Deuterium Alpha (FIDA) spectrometer and radially-scanning collimated neutron camera. In addition, strongly-driven chirping toroidicity-induced Alfv\'en eigenmodes are observed to cause fast-ion redistribution, as are sawteeth and large edge-localised modes. In each case, the modes affect fast ions in a region of real space governed by the eigenmode structure and principal toroidal mode number. Modelling using the global transport analysis code TRANSP, with \emph{ad hoc} anomalous diffusion introduced, reproduces the coarsest features of the affected fast-ion distribution in the presence of energetic-particle-driven modes, but the spectrally and spatially resolved FIDA measurements suggest that the distribution exhibits structure on a finer scale than is accounted for by this model.
\end{abstract}

\section{Introduction}\label{sec:intro}
The Mega-Amp Spherical Tokamak (MAST) is reliant on neutral beam heating to access high-performance operating regimes \cite{Akers2003, Buttery2004}. One of the world's largest spherical tokamaks, typical parameters of MAST are: major and minor radius $R=0.95\usk\metre$, $a=0.60\usk\metre$; plasma current $I_\mathrm{p}=400-900\usk\kilo\ampere$; toroidal field on axis $B_\mathrm{T}=0.40-0.58\usk\tesla$; core electron density and temperature $n_{\mathrm{e}0}=3\times10^{19}\usk\metre^{-3}$ and $T_{\mathrm{e}0}=1\usk\kilo\electronvolt$. The neutral beam injection (NBI) system presently installed on MAST consists of two positive ion sources, denoted south (SS) and south-west (SW), which are capable of accelerating deuterons to energies of $75\usk\kilo\electronvolt$. Each injector can deliver up to $2.5\usk\mega\watt$ of power when tuned to optimum perveance \cite{Gee2005}.

Fast ions in MAST are produced by ionization of beam neutrals as they propagate through the plasma. The relatively low magnetic field results in large fast-ion Larmor radii, which may be up to $10\usk\centi\metre$ for the highest energy beam ions. Once they are deposited, these ions interact with the background plasma, slowing down by electron drag at high energies and scattering off thermal ions at lower energies. Ultimately, the ions will either: thermalise, reaching equilibrium with the background plasma; undergo fusion reactions, producing a triton and proton, or $^3$He nucleus and neutron; become re-neutralised by charge exchange with beam or thermal neutrals; collisionally scatter onto unconfined orbits; or be lost from the plasma due to instabilities or static field perturbations. In order to maximise the electron and ion temperature and fraction of NBI-driven current in the plasma, and hence measures of performance such as plasma beta, it is desirable to confine the fast ions for as long as possible. Beam-driven current requires these ions to be confined at low plasma density, posing challenges for avoiding fast-particle-driven instabilities. Diagnosing the behaviour of fast ions in MAST, and the mechanisms for their redistribution within and possible loss from the plasma, is the objective of the present study.

At the ion temperatures that are achievable in MAST, thermonuclear fusion reactions make only a small contribution to the total neutron yield. Almost all ($>90\%$) of the fusion neutrons in MAST originate from beam-thermal and beam-beam reactions \cite{Turnyanskiy2013}. This makes the neutron emission from the plasma a useful diagnostic tool for tracking changes in the fast-ion population. In addition, the neutral beams themselves provide the potential to use active charge exchange spectroscopy to diagnose properties of the fast-ion distribution in both real and velocity space \cite{Heidbrink2004}. Both of these techniques are deployed on MAST, as outlined in the following section.

\section{Diagnostics}\label{sec:diag}
MAST is equipped with three principal fast-ion diagnostics. The first, a $^{235}$U fission chamber, provides a volume-integrated measurement of the total neutron emission from the plasma with a temporal resolution of $10\usk\micro\second$. Absolute calibration of the diagnostic was performed in 2008, and data are routinely available. More recently a collimated, radially-scanning neutron detector (known as the neutron camera, or NC) \cite{Cecconello2010} and a Fast-Ion Deuterium Alpha (FIDA) spectrometer \cite{Michael2013c} were installed. These instruments significantly extend the diagnostic capability by providing measurements of the fast-ion distribution with spatial resolution and, in the case of FIDA, resolution in velocity space. The temporal resolution of each of these diagnostics is approximately $1\usk\milli\second$.

\begin{figure*}[t]
\centering
\includegraphics[width=0.6\textwidth]{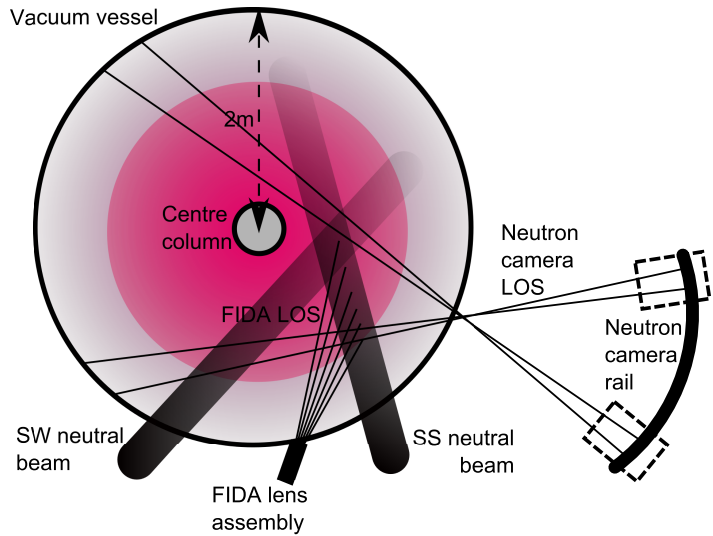}
\caption{Schematic plan view of the MAST vessel and neutral beam footprints, showing the neutron camera and FIDA diagnostic lines of sight. The neutron camera is shown at both ends of the rail, to indicate the full extent of the diagnostic's radial coverage. Also present, but omitted from the diagram for clarity, is a near-vertical FIDA view which looks down onto the SS neutral beam and a toroidally-displaced passive FIDA view which does not intersect either of the neutral beams.}\label{fig:LOS}.
\end{figure*}

The neutron camera simultaneously views four chords through the plasma, two of which lie on the machine midplane with different tangency radii, separated by $20\usk\centi\metre$, and two of which are angled downward such that at their tangency radii they are $20\usk\centi\metre$ below the midplane chords. The detector assembly is mounted on a curved rail so that it may rotate about a pivot point to scan the lines of sight across the plasma radius between discharges. Repeat discharges with the neutron camera in different positions may be performed to provide a neutron emissivity profile across the plasma radius; the quantity obtained is the emissivity integrated over the volume contained within each neutron camera channel's field of view. Precise characterisation of this field of view is therefore necessary in order to properly model the integrated emissivity and hence the total signal.

Concerning the FIDA diagnostic, two sets of views of the plasma are available. The first is a near-toroidal view from a lens mounted just above the vessel midplane. Lines of sight from this view are projected onto the path of one of MAST's neutral beam lines at major radii from $0.77\usk\metre$ to $1.43\usk\metre$, covering the full extent of the outboard and part of the inboard midplane. The second view is almost vertical, from a lens assembly mounted on the limb of one of the poloidal field coils and looking down on the path of the neutral beam. 11 lines of sight may be patched into the spectrometer at any given time; these may be all toroidal, all vertical, or a combination of the two. So-called `passive' lines of sight, which do not intersect a neutral beam line but which have otherwise similar viewing geometries, provide background subtraction of bremsstrahlung and passive FIDA light (which is produced by fast ions undergoing charge exchange with thermal neutrals at the plasma edge and subsequently emitting Balmer-alpha radiation).

Figure \ref{fig:LOS} depicts schematically the ranges of available viewing chords of the neutron camera and FIDA diagnostic. Since only data from the near-toroidal FIDA views are presented here, the geometry of the near-vertical views is not shown. As discussed in Reference \cite{Michael2013c}, the low transmission efficiency and hence poor signal-to-noise ratio of the near-vertical system makes these measurements unsuitable for studying the rapid changes associated with MHD modes. Further details of the geometry of each of the diagnostics are provided in the literature \cite{Cecconello2010, Michael2013c}.

\begin{figure}[h]
\centering
\includegraphics[width=0.6\textwidth]{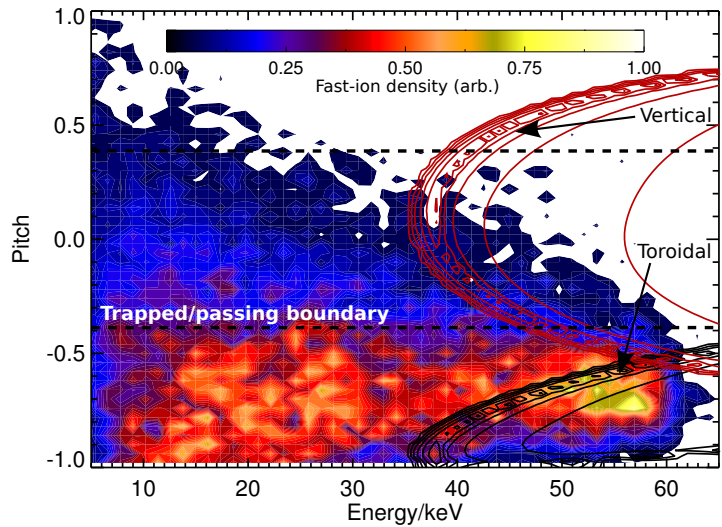}
\caption{Typical synthetic fast-ion distribution $f(E,p)$ at $R=1.0\usk\metre$, $Z=0.0\usk\metre$ from TRANSP for a 2-beam, double-null diverted MAST discharge, with FIDA weight functions for the toroidally and vertically-viewing systems overplotted. The full injection energy of both beams during this discharge was $59\usk\kilo\electronvolt$. Black contours show the sensitivity of the toroidally-viewing system at $R=1.00\usk\metre$, $\lambda=660.2\usk\nano\metre$. Red contours show the sensitivity of the vertically-viewing system at the same radius, at $\lambda=652.0\usk\nano\metre$. The approximate location of the trapped/passing boundary is also shown for reference.}
\label{fig:weight}
\end{figure}

In addition to the spatial coverage and resolution of each of the diagnostics, important consideration must be given to their sensitivity in velocity space. The DD fusion reactions which provide the neutron camera signal have a relatively simple energy dependence; the reaction cross-section is a steeply-rising function of centre-of-mass energy over the energy range relevant to beam ions in MAST, i.e. $1\usk\kilo\electronvolt - 75\usk\kilo\electronvolt$ \cite{Bosch1992}. FIDA measurements on the other hand have a non-trivial dependence on fast-ion energy and pitch ($p=v_\parallel/v$); their sensitivity in velocity space may be represented by a \emph{weight function} parametrised by the angles between the line of sight and the neutral beam, and the line of sight and magnetic field. An example of such a weight function is shown in Figure \ref{fig:weight}. The FIDA signal observed at a particular spatial location in a given wavelength range is a product of the weight function at that location and wavelength, and the local fast-ion density in the region of velocity space bounded by this weight function (which forms an arc in energy/pitch space as seen in Figure \ref{fig:weight}). Since the FIDA measurements are spectrally-resolved, an additional degree of freedom in interpretation of the data is provided by the choice of wavelength. Choosing a particular wavelength corresponds to choosing a minimum fast-ion energy to which the measurement is sensitive; for the toroidal FIDA system, selecting a higher wavelength allows one to translate the weight function to the right of Figure \ref{fig:weight}, i.e. to higher energy. Observing D$\alpha$ emission at a wavelength of $660.2\usk\nano\metre$, chosen for the toroidal weight function plotted in black in Figure \ref{fig:weight}, corresponds to observing only those ions with energies of at least 36.6 keV (although this is extended to slightly lower energies in practice by the finite spectral resolution of the diagnostic).

As Figure \ref{fig:weight} shows, the highest energy ions in the core of MAST plasmas lie mainly in the passing region of velocity space where the high parallel component of their velocity allows these particles to traverse the full poloidal extent of a flux surface during their orbits. As they slow down by electron drag, the fast ions start to undergo an increasing amount of pitch-angle scattering off thermal ions; this causes some ions to cross the trapped/passing boundary, at which point their orbits are transformed into trapped banana orbits. The toroidally-viewing FIDA system from which results are presented in this paper is mostly sensitive to passing particles with large $|p|$, particularly near the core of the plasma. Similarly, the neutron emission, being derived mostly from high energy fast ions reacting with thermal ions, is weighted to the high energy part of velocity space in which a large majority of the fast ions lie on passing orbits. We therefore expect changes in the fast-ion density in this region of velocity space to be apparent as changes of a similar nature in both the toroidal FIDA and the neutron camera signals. This expectation of consistency between the FIDA and NC signals lies behind the power of the multi-diagnostic analysis presented here. There are however situations in which information may be obtained from differences between the behaviour of the FIDA and NC signals; for example, an instability which causes pitch-angle scattering of fast ions from the passing to the trapped region of velocity space will cause the FIDA signal to be depleted, but as long as this process does not change the energy of the ions then the neutron camera signal should be largely unaffected.

We now present the results of analysis of FIDA and NC data from MAST discharges exhibiting frequency-chirping energetic particle modes, including both fishbones and toroidicity-induced Alfv\'en eigenmodes (TAEs). Note that in the sections which follow, figures presenting FIDA and NC data which show time traces from multiple channels in a single plot have had each trace independently scaled to prevent multiple lines overlaying each other and to therefore aid clarity. Where this applies, the caption of the relevant figure includes a \emph{scaling factor} (SF) for each time trace; the data have been multiplied by this factor, thus preserving the origin of the vertical axis and the relative magnitude of fluctuations in data from each channel.

\section{Chirping modes}\label{sec:chirp}
\subsection{Chirping modes observed in MAST plasmas}
At high NBI power and plasma current, many MAST discharges include periods of chirping mode activity. Commonly, these periods consist of a phase shortly after beam switch-on in which chirping TAEs are active, with frequencies of around $100\usk\kilo\hertz$, followed by a transition to a series of quasi-periodic fishbones at lower frequency, around $40\usk\kilo\hertz$.

\begin{figure}[ht]
\begin{centering}
\includegraphics[width=0.6\textwidth]{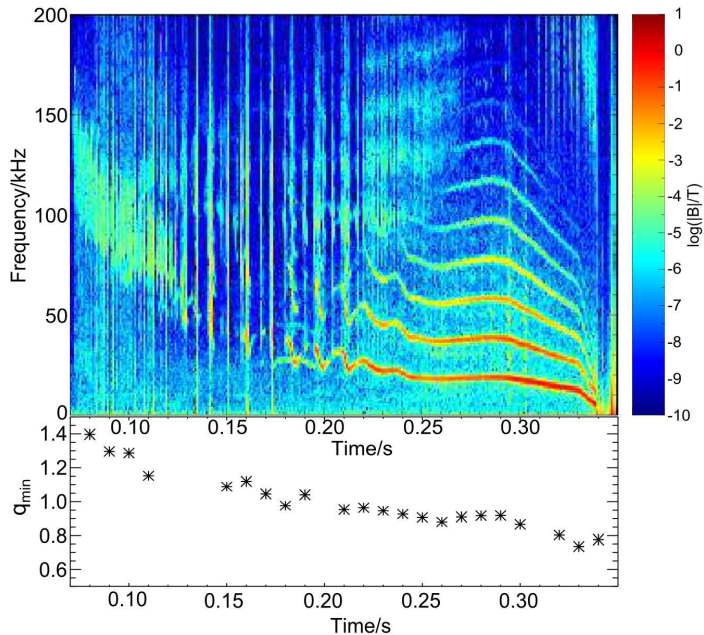}
\caption{Magnetic spectrogram of a typical beam-heated MAST discharge (\#29994) from a Mirnov pick-up coil located on the outboard midplane. Beam switch-on occurs at $0.07\usk\second$. Initial activity at TAE frequencies ($\sim100\usk\kilo\hertz$) eventually gives rise to well-separated chirping modes with central frequencies evolving from $50\usk\kilo\hertz$ down to $30\usk\kilo\hertz$ (i.e. fishbones). These develop into the long-lived mode at around $0.22\usk\second$. The LLM saturates and locks to the plasma rotation, dropping rapidly in frequency shortly before $0.34\usk\second$. Higher harmonics of the LLM, up to $n=8$, are also observed. The plasma disrupts at $0.35\usk\second$. Also plotted, in the second panel, is the minimum value of the safety factor $q_\mathrm{min}$ determined from MSE-constrained EFIT.}\label{fig:spectro}
\end{centering}
\end{figure}

It is clear from examination of spectrograms such as that shown in Figure \ref{fig:spectro} that rather than being entirely distinct modes, the fishbones and TAEs represent different stages in a continual evolution. The initial frequency of the modes evolves gradually downward over time, and the modes become less `Alfv\'en-like' and more `fishbone-like' in nature; eventually these fishbones evolve into the saturated long-lived internal kink mode (LLM). The key factor which determines the evolution of MHD activity throughout the discharge is the q-profile; as the minimum value of the safety factor, $q_\mathrm{min}$, evolves downward throughout the shot, the plasma becomes unstable first to chirping then to steady-state internal kink modes. The evolution of $q_\mathrm{min}$ is also shown in Figure \ref{fig:spectro}.

\begin{figure}[h]
	\begin{subfigure}[t]{0.5\textwidth}
	\centering
	\includegraphics[width=\textwidth]{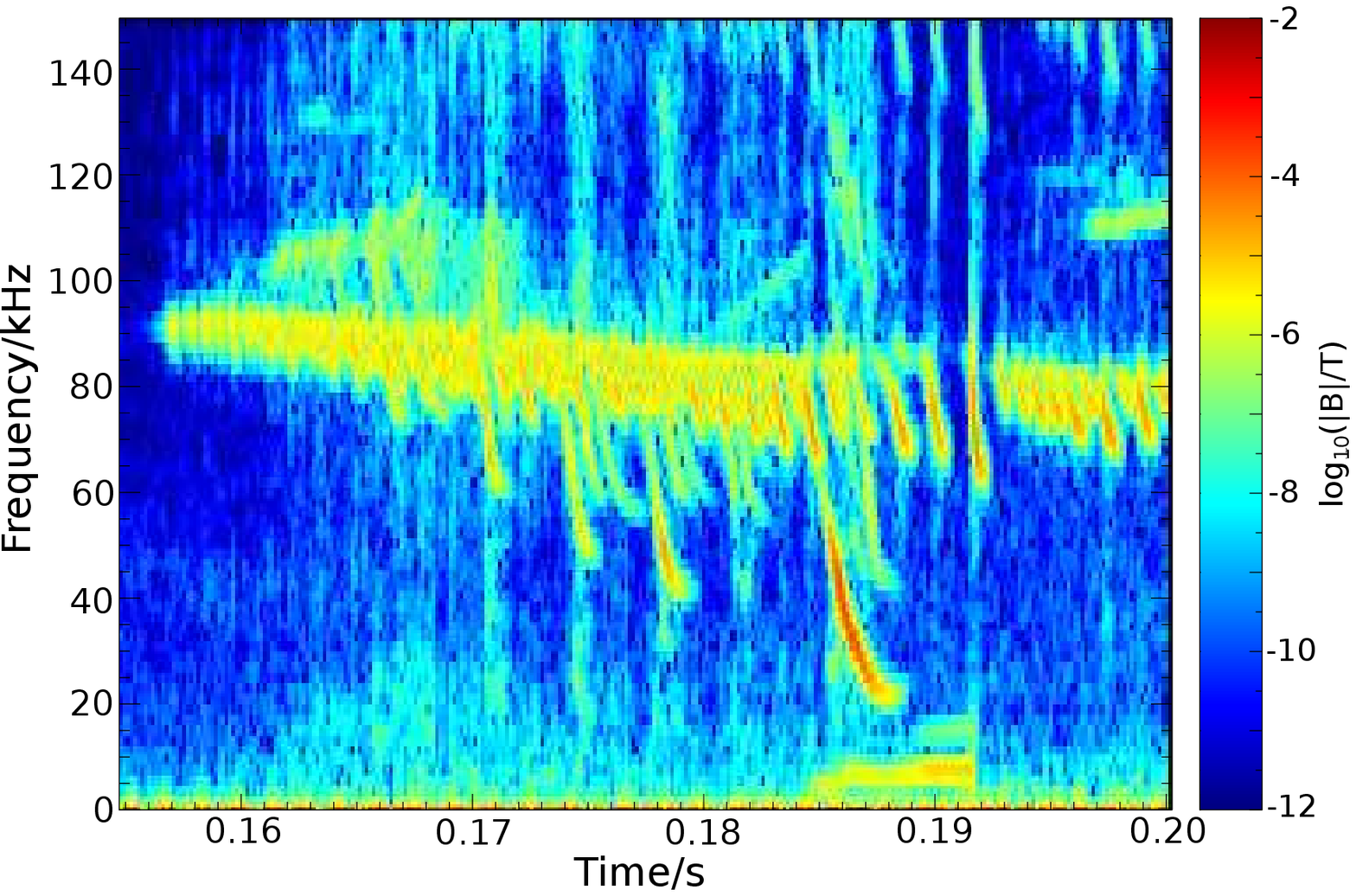}
	\caption{}
	\label{fig:taespectro}
	\end{subfigure}
	\begin{subfigure}[t]{0.5\textwidth}
	\centering
	\includegraphics[width=\textwidth]{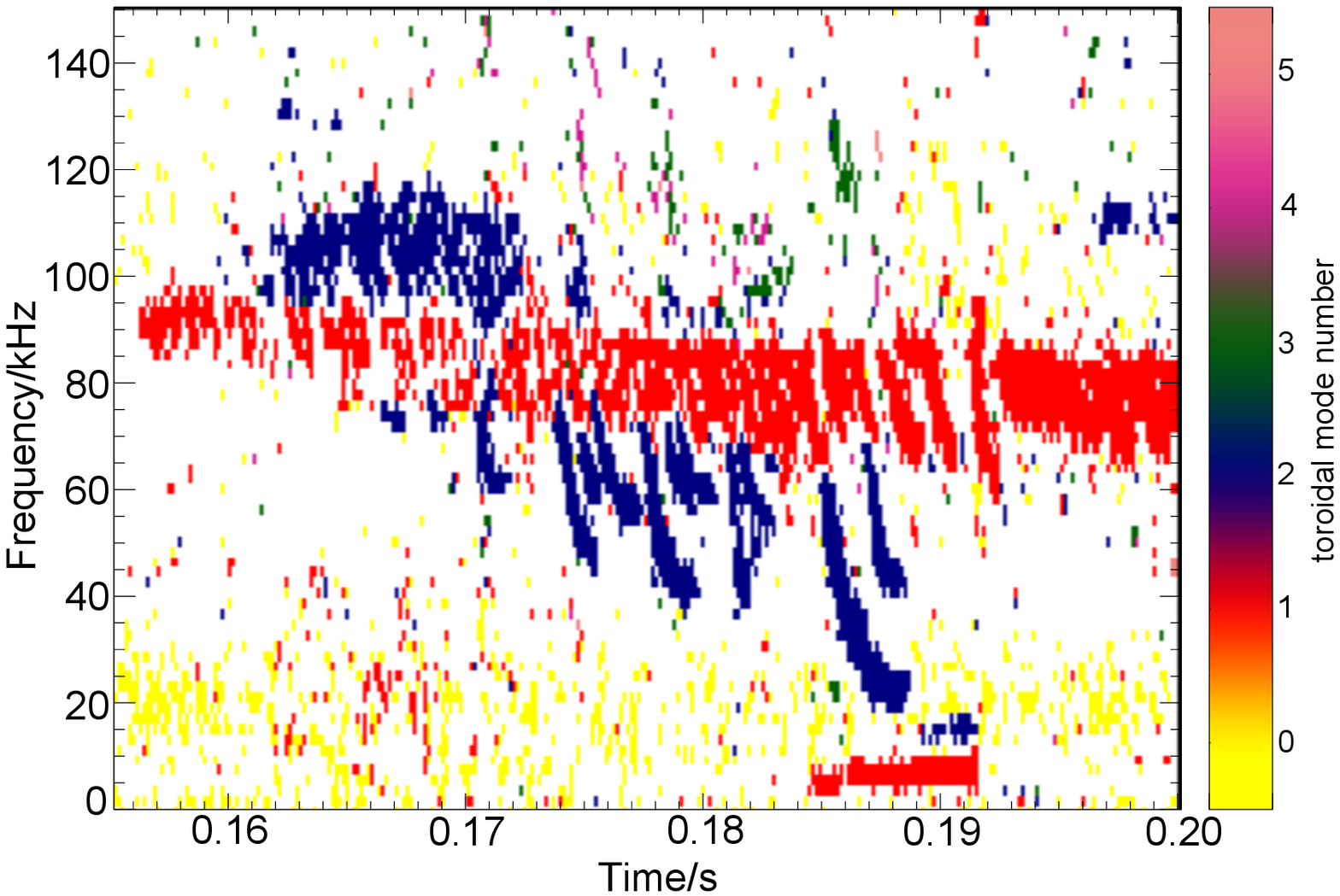}
	\caption{}
	\label{fig:taenmode}
	\end{subfigure}
\caption{(a) Spectrogram of poloidal magnetic field perturbation amplitude and (b) toroidal mode number analysis of poloidal field perturbations during MAST shot \#29902, derived from an array of Mirnov pickup coils on the outboard midplane.}\label{fig:taechirp}
\end{figure}

The two factors which primarily dictate the strength and nature of the chirping mode activity in MAST are the plasma density and beam power. At high beam power and low density, the long fast-ion slowing down time ($>30\usk\milli\second$) and consequent large population of particles at high energies results in a high fast-ion beta. This results in strongly driven modes exhibiting nonlinear phenomena such as explosive growth in amplitude and strong frequency chirping. In one recent MAST discharge, chirping TAE modes were observed to sweep down to frequencies as low as $20\usk\kilo\hertz$; the NBI power was $2\usk\mega\watt$ and the primary energy of the beam ions was $70\kilo\electronvolt$. Figure \ref{fig:taechirp} shows the amplitudes and mode numbers of TAEs present at the time of this event; chirping TAEs with toroidal mode number $n=1$ are present throughout the interval shown, but the largest events with the strongest frequency chirp have $n=2$. It is noteworthy that in a near-identical shot with lower beam voltage and power ($60\usk\kilo\volt$, $1.5\usk\mega\watt$), fewer $n=2$ chirping modes were observed although the frequencies and evolution of the $n=1$ train of TAE activity and the clump of $n=2$ precursor modes, starting just after $0.16\usk\second$ and persisting until the first $n=2$ chirp, were very similar. The largest event, at approximately $0.185\usk\second$ in Figure \ref{fig:taechirp}, was entirely absent from the lower power discharge. Phenomenologically, the explosive nonlinear evolution of the $n=2$ TAE is reminiscent of the TAE avalanches reported on NSTX \cite{Podesta2011}; the overlapping frequencies and similar evolution of the $n=1,2$ modes support the idea of wave-wave coupling.

\subsection{Fishbones}
\begin{figure}[h]
	\begin{subfigure}[t]{0.48\textwidth}
	\centering
	\includegraphics[width=\textwidth]{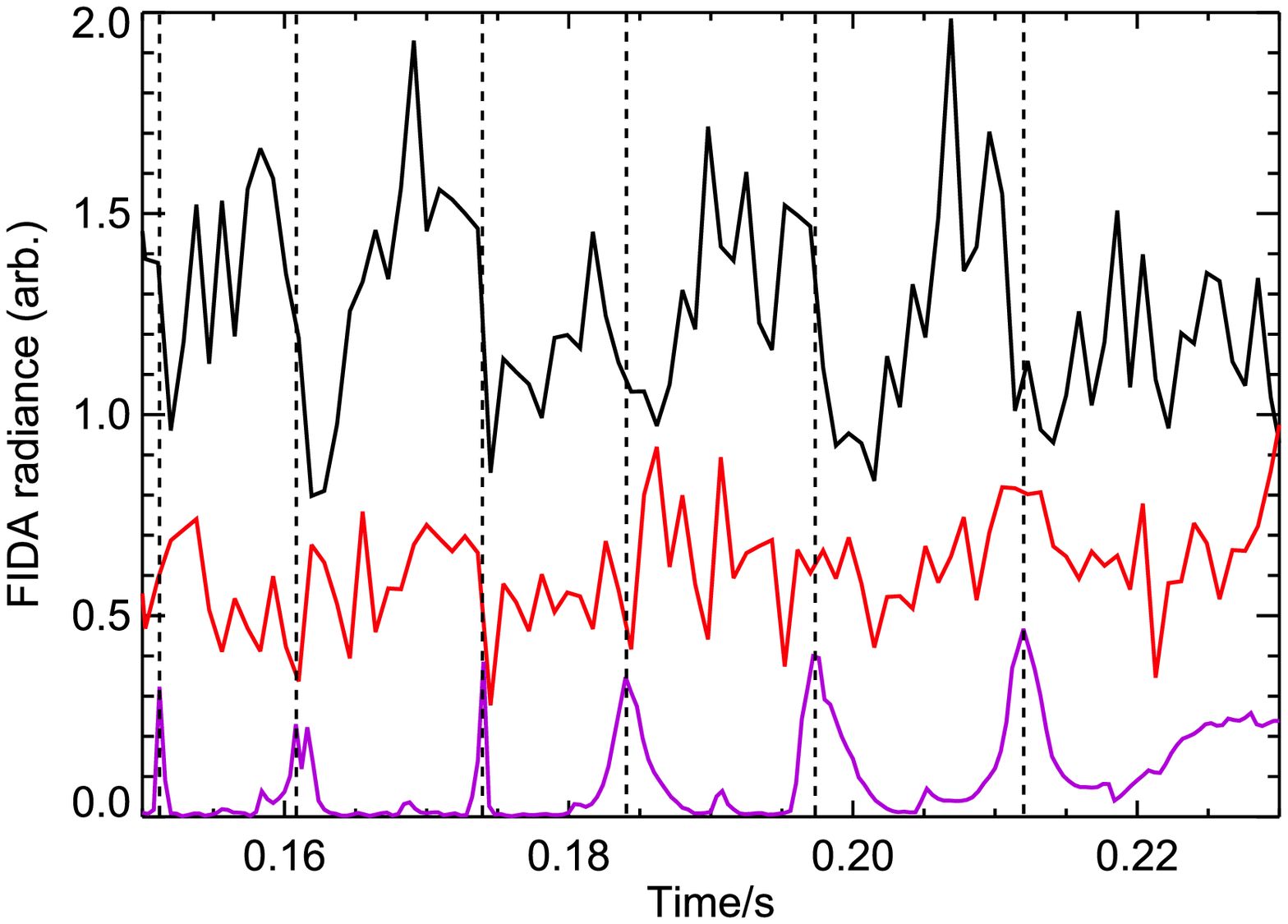}
	\caption{}
	\label{fig:fbtrace}
	\end{subfigure}
	\begin{subfigure}[t]{0.48\textwidth}
	\centering
	\includegraphics[width=\textwidth]{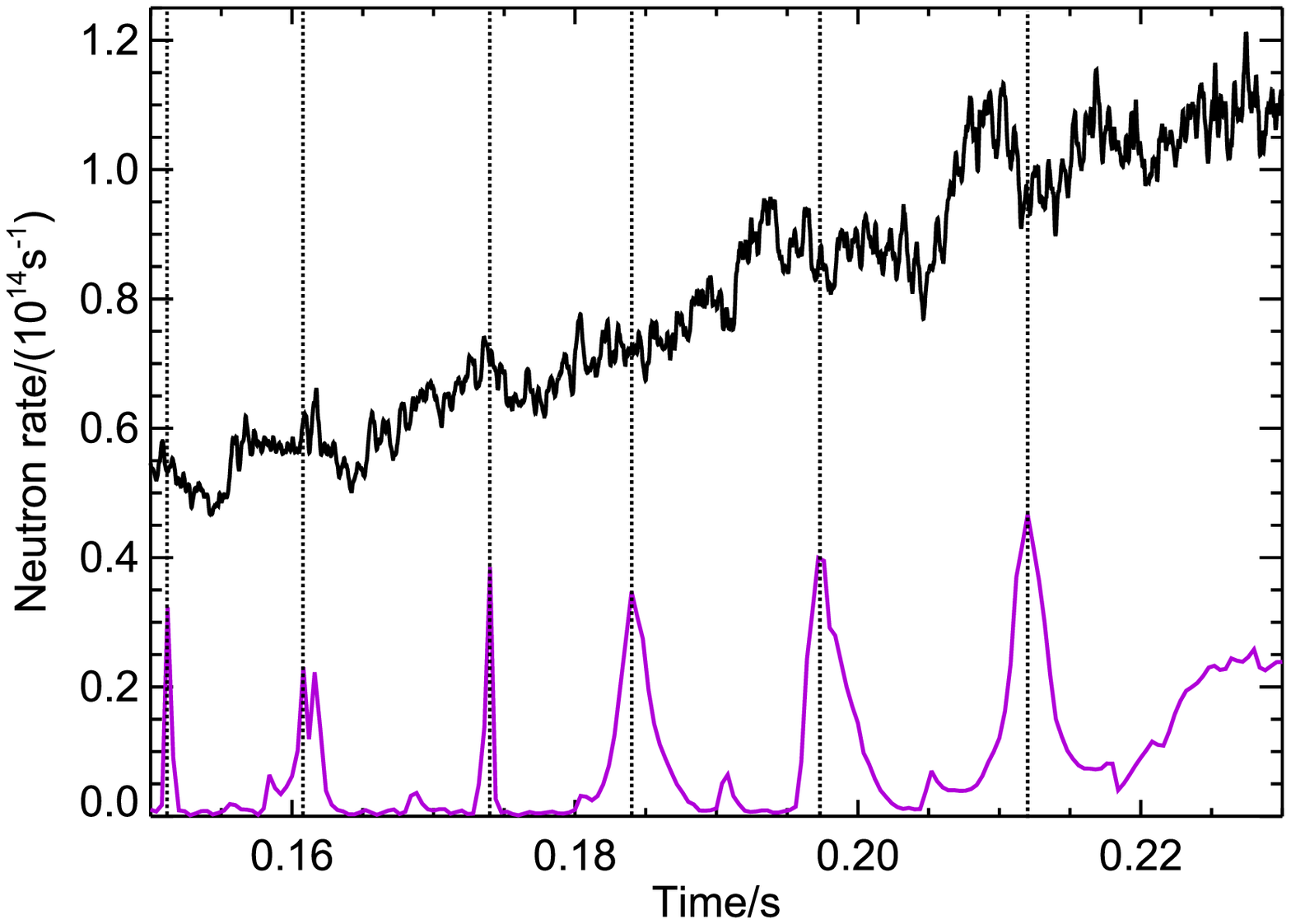}
	\caption{}
	\label{fig:anufb}
	\end{subfigure}
\caption{(a) FIDA signal ($\lambda=661.1\usk\nano\metre$) at two different radii during a period of MAST shot \#29994 in which fishbones were active. Black trace: $R=1.07\usk\metre$; SF = 1.3. Red trace: $R=1.27\usk\metre$; SF = 1.1. The purple trace shows the RMS amplitude of the outboard midplane Mirnov coil signal; clear, rapid drops in the core FIDA signal (black) are observed to be correlated with the large fishbones, as well as with the much smaller events which occur between the main bursts. (b) Global neutron rate from the fission chamber during the same period. The RMS amplitude of the outboard midplane Mirnov coil signal is also shown for reference.}\label{fig:fb}
\end{figure}

Examining FIDA data from shot \#29994 in Figure \ref{fig:fbtrace}, we see clear `cycles' of drops in signal from the channel which views the core of the plasma, associated with individual fishbones (the purple curve shows the RMS amplitude of the Mirnov coil signal, which is a measure of $\dot{B}_\theta$). The drops in signal at the outer radius are substantially weaker, indicating localised depletion of the fast-ion density. The line of sight/neutral beam intersection radii chosen for the time traces of FIDA data in Figure \ref{fig:fbtrace} are $1.07\usk\metre$ for the core channel (black curve) and $1.27\usk\metre$ for the outer channel (red curve). These findings are largely consistent with those reported in earlier work \cite{Jones2013}, although the effect of fishbones on FIDA signal at larger radii does not seem to be as great in the present case. One possible reason for this is that the magnetic perturbations are weaker by approximately a third than in previously reported cases. Additionally, little confidence could be placed in data from radii outside approximately $1.25\usk\metre$ in the earlier work due to a poor choice of lines of sight for background subtraction; this has subsequently been improved. Previous results from the neutron camera \cite{Cecconello2012} showed significant drops in the neutron rate for the core midplane and lower channels, at $R=1.00\usk\metre,\:Z=0.00\usk\metre$ and $R=1.00\usk\metre,\:Z=-0.20\usk\metre$ respectively. Although neutron camera data are unavailable for shot \#29994, Figure \ref{fig:anufb} shows the global neutron rate from the fission chamber over the same period featured in Figure \ref{fig:fbtrace}. Drops are associated with most of the chirping modes, with the notable exception of the fishbone at $0.184\usk\second$, but these drops are much weaker in magnitude than those in the spatially and energetically-localised core FIDA signal in Figure \ref{fig:fbtrace}. This finding supports the idea that fast ions are mostly redistributed within, rather than lost from, the plasma; since the neutron emission results predominantly from beam-thermal fusion reactions, this redistribution has only a small effect on the total neutron yield.

\begin{figure}[h]
	\begin{subfigure}[t]{0.48\textwidth}
	\centering
	\includegraphics[width=\textwidth]{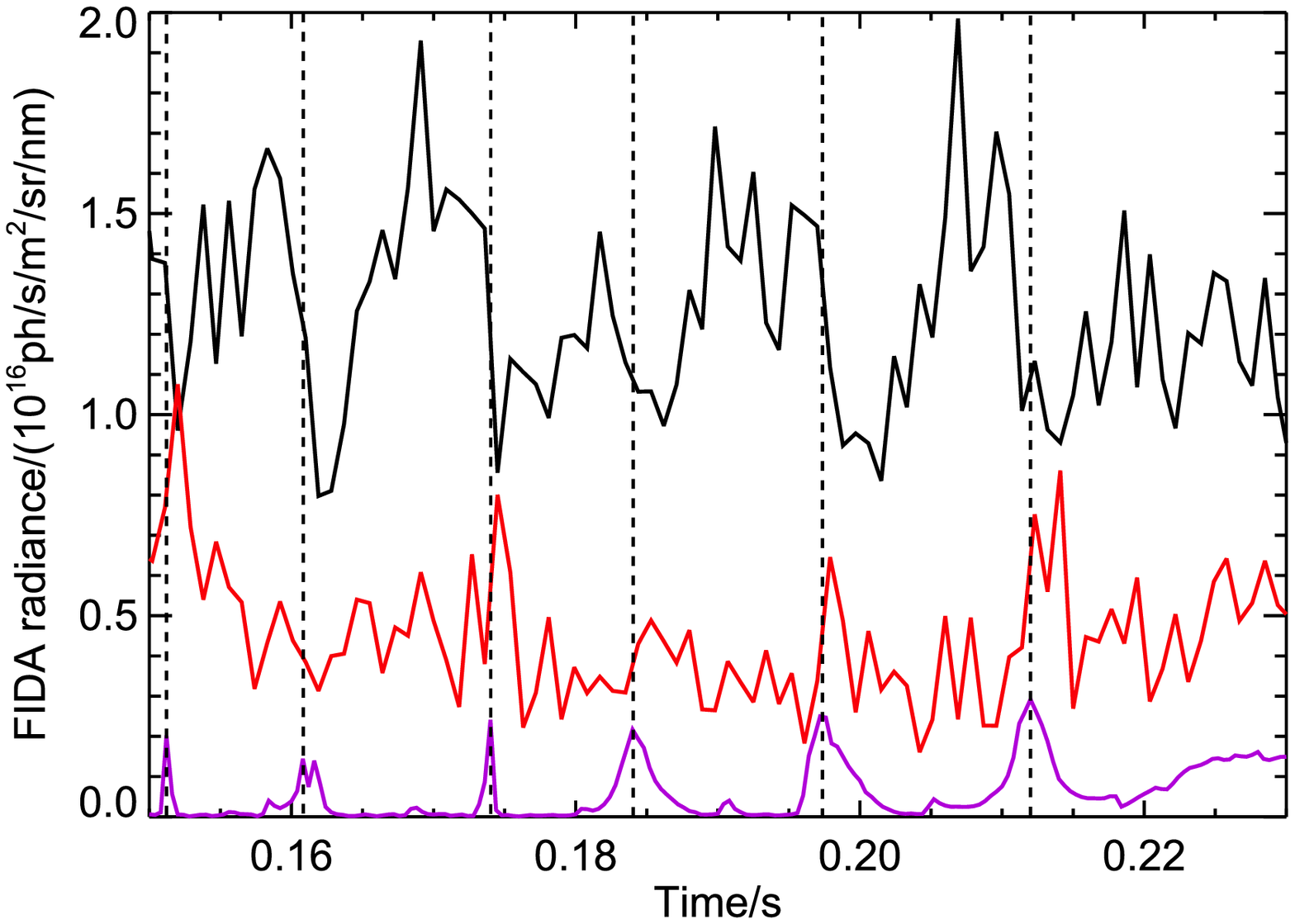}
	\caption{}
	\label{fig:fbtrace2}
	\end{subfigure}
	\begin{subfigure}[t]{0.48\textwidth}
	\centering
	\includegraphics[width=\textwidth]{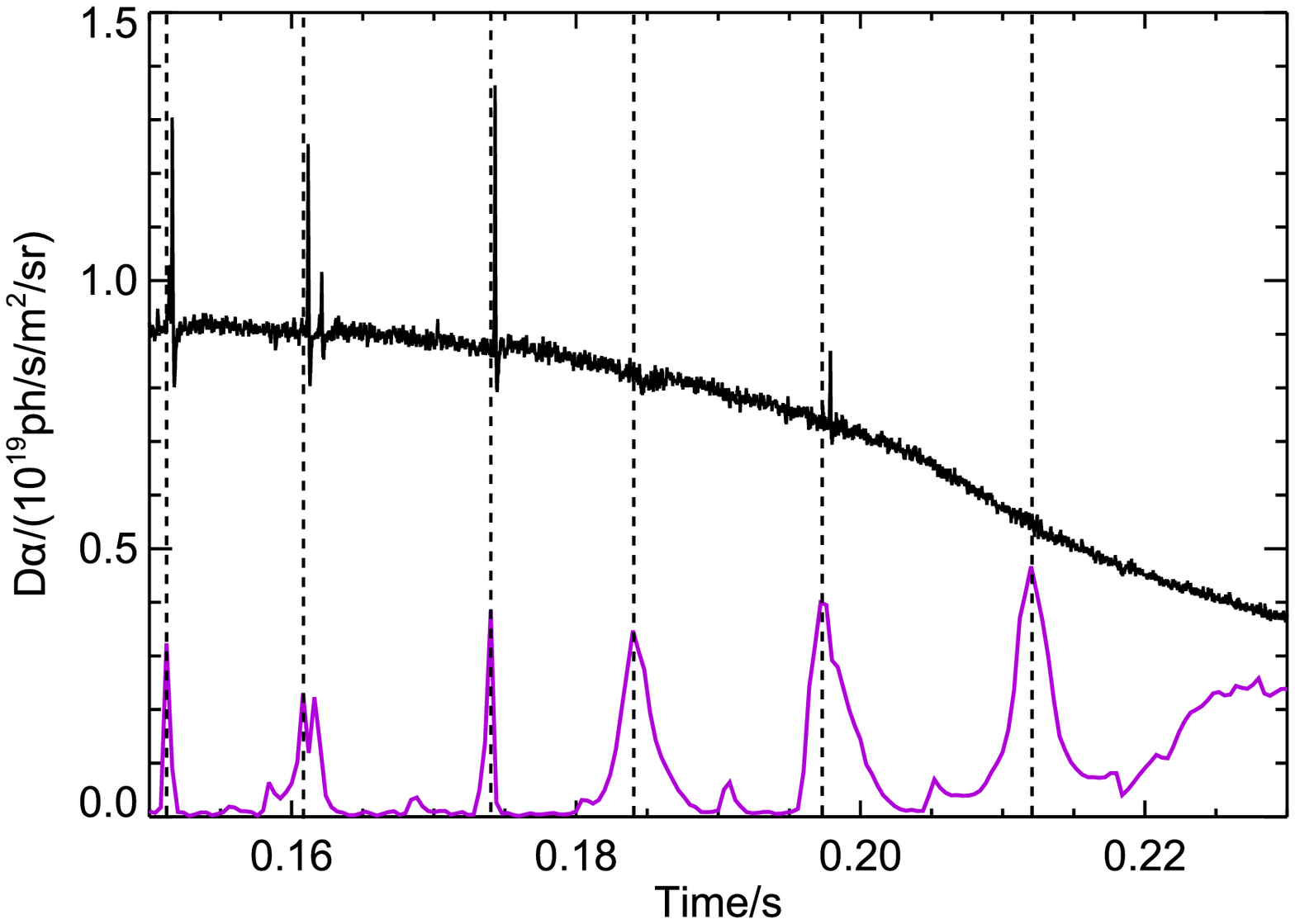}
	\caption{}
	\label{fig:adafb}
	\end{subfigure}
\caption{(a) FIDA signal ($\lambda=661.1\usk\nano\metre$) at two different radii during a period of MAST shot \#29994 in which fishbones were active. Black trace: $R=1.07\usk\metre$; SF = 1.25. Red trace: $R=1.33\usk\metre$; SF = 0.9. The purple trace shows the RMS amplitude of the outboard midplane Mirnov coil signal. (b) D$\alpha$ light from the plasma boundary during the same period. The RMS amplitude of the outboard midplane Mirnov coil signal is also shown for reference.}\label{fig:fb2}
\end{figure}

Regarding the question of the region to which fast ions are redistributed by the fishbones, some indication of an outward radial flux is provided by Figure \ref{fig:fbtrace2}. This shows the fishbone-induced changes in the core signal at $R=1.07\usk\metre$, as in Figure \ref{fig:fbtrace}, but also reveals transient increases in the signal at $R=1.33\usk\metre$ (with the exception of the burst at $0.161\usk\second$). The temporal evolution is consistent with the identification of these increases as resulting from the radial displacement of fast ions from the core of the plasma due to the fishbone, but the fact that these increases are relatively spatially localised and are not observed for all the bursts in magnetic activity suggests that the redistributed fast ions also undergo pitch-angle scattering which removes them from the part of velocity space to which most of the FIDA views are sensitive. One further time trace which hints at the ultimate fate of the fast ions redistributed from the core is provided by the edge D$\alpha$ signal in Figure \ref{fig:adafb}. The first three bursts in the Mirnov coil signal have a very rapid chirp with several harmonics of the magnetic activity overlapping at frequencies typical of TAEs, at around $100\usk\kilo\hertz$. Each of these bursts is observed to cause a large spike in the edge D$\alpha$ emission, whereas no such spike is observed at the time of the three later events which evolve more slowly in amplitude and have none of the TAE characteristics of the first three bursts. Note also that TAE bursts earlier in this discharge, which had no apparent low-frequency MHD component, were also observed to cause spikes in the D$\alpha$ signal when the bursts were well-separated with a relatively large amplitude. It is posited that these spikes in the D$\alpha$ light from the plasma boundary arise from reneutralized fast ions which have undergone charge exchange with thermal neutrals at the edge of the plasma - in other words they correspond to passive FIDA light from lost fast ions with near-zero velocity along the line of sight (such that the FIDA emission is not Doppler shifted). As expected based on this interpretation, similar spikes are observed in both passive (non-beam-viewing) and active (beam-viewing) FIDA channels, and these spikes are indeed significantly larger for the TAE-fishbones (first three events) than the `classic' internal kink-fishbones (latter three events). Furthermore, the only `classic' fishbone for which a small spike is observed in the edge D$\alpha$ trace in Figure \ref{fig:adafb} is also seen to cause the largest perturbation to the passive FIDA signal at the edge. These observations indicate that fishbones cause a displacement of fast ions radially outward, probably with associated pitch-angle scattering, while only the chirping modes with a significant TAE-like nature actually cause losses of fast ions from the plasma boundary. Note that in terms of a fraction of the fast-ion population, the modest effect on global neutron rate implies that these losses are rather small. Furthermore, consistent with previous observations \cite{Jones2013}, it appears that there is a cut-off energy above which fast ions are relatively unaffected by these modes; the drops in core signal and spikes in passive edge signal are negligible beyond $\lambda=661.6\usk\nano\metre$, which corresponds to a minimum fast-ion energy of $66\usk\kilo\electronvolt$. The beam injection energy in this shot was $70\usk\kilo\electronvolt$.

\begin{figure}[h]
\centering
\includegraphics[width=0.505\textwidth]{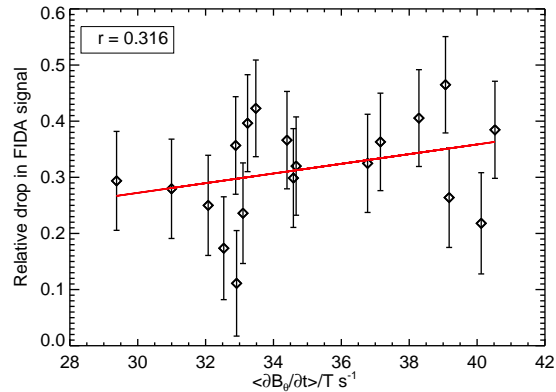}
\caption{Correlation between the relative change in FIDA signal at $R=0.98\usk\metre$, $\lambda=660.6\usk\nano\metre$, and the peak RMS time derivative of the poloidal magnetic field perturbation $\langle\dot{B}_\theta\rangle_\mathrm{RMS}$ as measured by a Mirnov coil on the outboard midplane. The linear regression coefficient, $r=0.316$, supports correlation with 80\% confidence. The reduced chi-square of the fit is 0.969.}\label{fig:corrplot}
\end{figure}

In previous work, a strong correlation was found between the maximum time derivative of the poloidal magnetic field perturbation and the relative change in FIDA signal associated with fishbones \cite{Jones2013}. Although the recent experimental campaign on MAST did not feature a series of repeated discharges with such well-separated fishbones, one candidate scenario was identified in which the fishbones were numerous enough and the discharges repeatable enough to provide a reasonable database of fishbone events. For the set of shots \#29207 - \#29210, fishbones occurred between $0.220\usk\second$ and $0.260\usk\second$, with a typical inter-fishbone period of $8-9\usk\milli\second$. The relative change in FIDA signal resulting from these fishbones is plotted against the peak RMS time derivative of the poloidal field perturbation in Figure \ref{fig:corrplot}. The result, with $r=0.316$, lends only weak support to the existence of a correlation. This value of the regression coefficient implies a significance of 80\%. The weighted, reduced chi-square value for the fit is 0.969. A possible reason for the much weaker correlation in this case is the presence of almost continuous magnetic perturbations between the fishbone events, which will affect the FIDA signal in a way which this analysis, which considers only the peak Mirnov signal amplitude, does not take into account. In addition, the more closely spaced fishbones in these shots require binning the FIDA data over narrower intervals, resulting in a dataset which is strongly affected by noise. Worth noting is the fact that as in the case of the fishbones reported in Reference \cite{Jones2013}, no correlation was found between the change in FIDA signal and the true \emph{amplitude} of the perturbation ($r\sim0.01$).

\begin{figure}[h]
\centering
\includegraphics[width=0.505\textwidth]{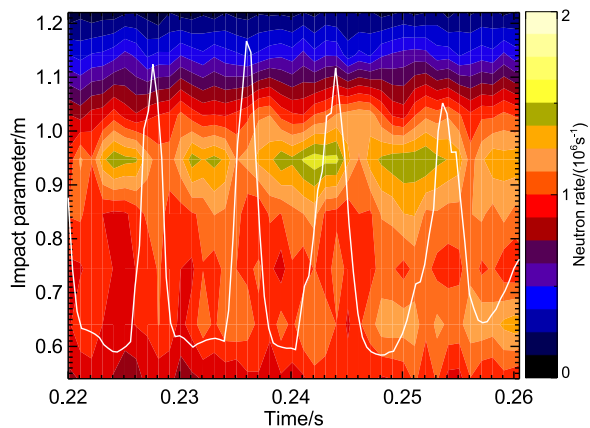}
\caption{Radial profile of neutron camera count rates during a period with quasi-periodic fishbones observed in MAST shot \#29207 and subsequent repeated discharges. The fishbones were time-aligned according to the magnetics signal from the outboard midplane Mirnov coil, and this time offset was applied to each of the NC time traces to generate this `composite' radial profile as a function of time. Overplotted in white is the RMS Mirnov coil time trace from shot \#29207.}\label{fig:NCscan}
\end{figure}

Extending the analysis of the effects of fishbones at different radii, the same series of shots used for the analysis summarised in Figure \ref{fig:corrplot} provided an opportunity to scan the neutron camera in major radius. With two radial points per pulse, the series of four pulses provided an eight-point radial scan. Figure \ref{fig:NCscan} shows the combined result during the period of each of these shots when quasi-periodic fishbones were observed. The fishbones were `time-aligned' with respect to the Mirnov coil traces, so that the fishbones from each shot overlay those of the other shots in the series as closely as possible. The same offset required to achieve this alignment was then applied to the neutron camera timebase, to build a picture of the behaviour of the NC signal at different radii throughout the fishbone cycle. Clear cycles of a build-up in neutron rate, followed by a rapid drop once the mode grows, are seen in particular at core radii, between $R=0.9\usk\metre$ and $1.0\usk\metre$. This corresponds to the position of the magnetic axis in MAST.

\subsection{Toroidicity-induced Alfv\'en Eigenmodes}
\begin{figure}[h]
	\begin{subfigure}[t]{0.4935\textwidth}
	\centering
	\includegraphics[width=\textwidth]{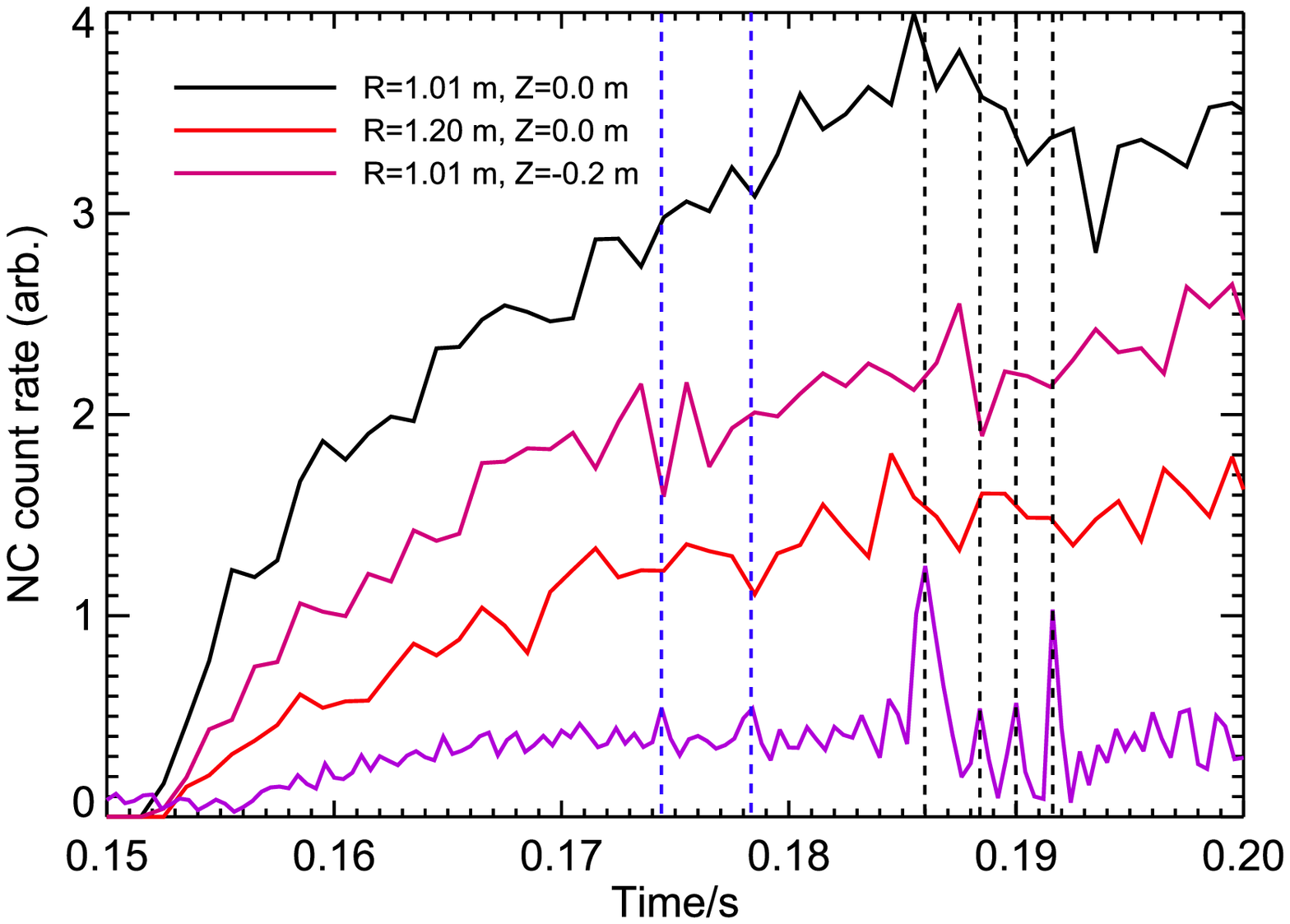}
	\caption{}
	\label{fig:nctae}
	\end{subfigure}
	\begin{subfigure}[t]{0.5065\textwidth}
	\centering
	\includegraphics[width=\textwidth]{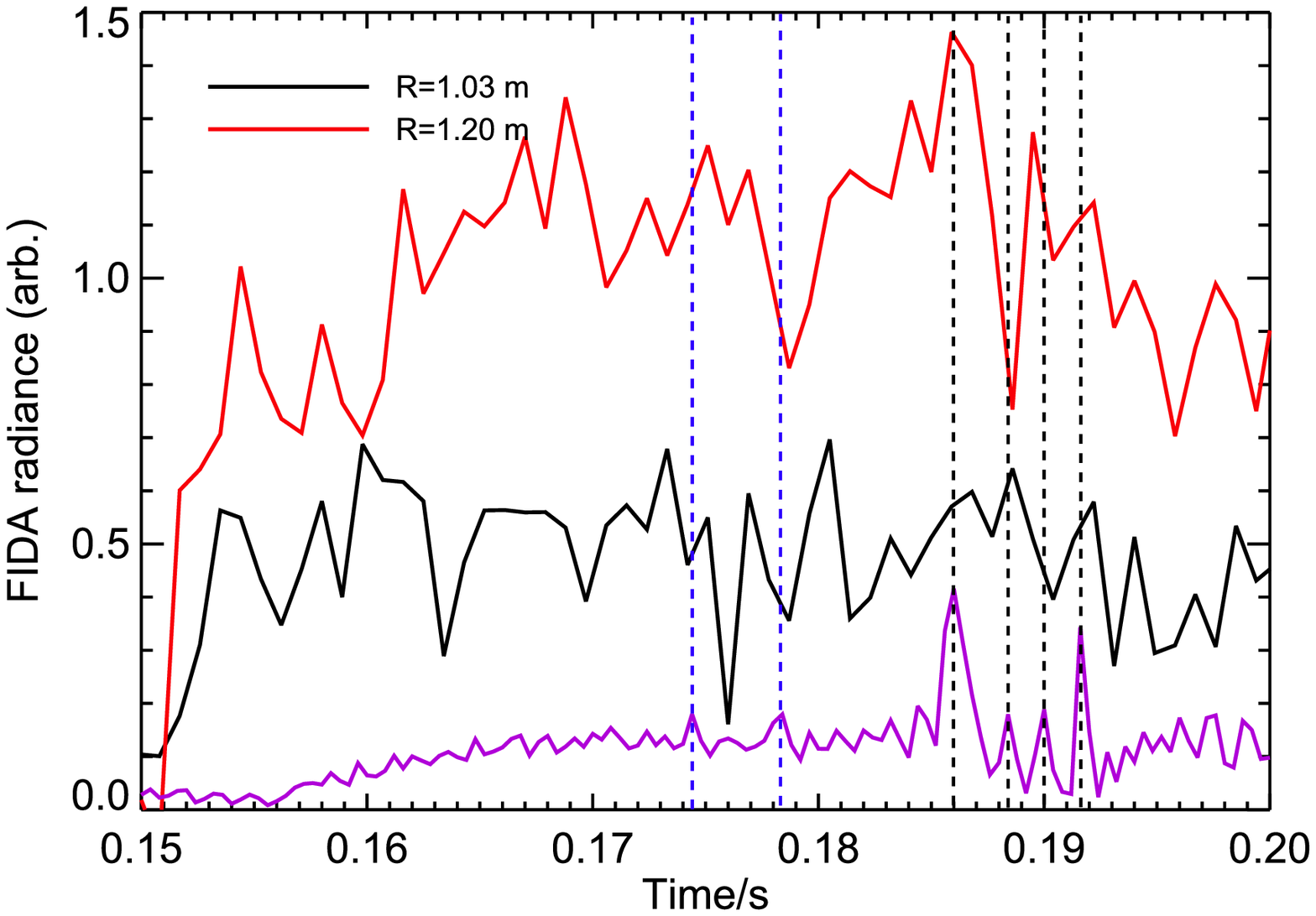}
	\caption{}
	\label{fig:fidatae}
	\end{subfigure}
\caption{(a) Neutron camera data from MAST discharge \#29902 around the time of the chirping TAE modes seen in Figure \ref{fig:taechirp}. Black (SF = 1.0) and red (SF = 0.8) traces show data from the core and outer midplane channels, while the magenta (SF = 1.1) trace is obtained from the channel viewing below the midplane, at the same major radius as the core midplane channel. (b) FIDA data ($\lambda=661.2\usk\nano\metre$) during the same period. The black trace shows the signal from the core channel, and the red trace shows the signal from further out in major radius; line of sight/neutral beam intersection radii are $1.03\usk\metre$ (black; SF = 1.5) and $1.20\usk\metre$ (red; SF = 1.35). The RMS amplitude of the Mirnov coil signal is also shown in each panel for reference.}\label{fig:tae}
\end{figure}

Turning our attention to the chirping TAE modes identified in Figure \ref{fig:taechirp}, data from both the FIDA diagnostic and the NC are available. The neutron camera signal from the outer midplane channel, shown in red in Figure \ref{fig:nctae}, exhibits a drop associated with the large $n=2$ chirping event at $0.186\usk\second$. The perturbation to the signal in the core midplane channel (black) is much smaller than the subsequent drop associated with the three large, clearly-separated $n=1$ TAE bursts between $0.188\usk\second$ and $0.192\usk\second$. The outer midplane and diagonally-oriented chords by comparison see a much smaller change in signal associated with these $n=1$ bursts.
It is particularly interesting that while the outer midplane view sees small drops in signal associated with well-separated TAE bursts, the below-midplane view actually detects an \emph{increase} in integrated neutron emissivity. This could indicate a poloidal asymmetry in the fast-ion redistribution caused by these modes. Blue, dashed vertical lines in Figure \ref{fig:nctae} indicate the times at which other large bursts of $n=2$ TAE activity occur; the first of these is seen to cause a significant perturbation to the signal from the diagonally-oriented channel (magenta), while the second does not cause an appreciable change in count rate in any of the channels. These results suggest that each TAE burst causes selective redistribution of fast ions to and from regions of real or velocity space which are not necessarily the same as those which drive the mode unstable. This is in contrast to the case of the fishbones, where the `limit cycle' behaviour in the core FIDA signal suggests that those fast ions which drive the fishbone unstable are also those which are most strongly affected by the mode. In addition, no spikes in the edge D$\alpha$ or passive FIDA channels are observed in the case of any of these TAEs, suggesting that the fast ions remain confined.

FIDA data from the same period are shown in Figure \ref{fig:fidatae}. As in the case of the neutron camera data, the effect of the large $n=2$ chirp is much greater on the signal at $R=1.20\usk\metre$ than on the core signal at $R=1.03\usk\metre$. The subsequent $n=1$ chirping modes also affect the core FIDA signal strongly, as they affected the core NC signal. One point to note, which indicates consistency between the FIDA and NC observations, is that the second large $n=1$ chirp at $0.190\usk\second$ causes the core signal (black trace) in both Figures \ref{fig:nctae} and \ref{fig:fidatae} to recover slightly, whereas the preceding and subsequent $n=1$ chirps cause the signal at this radius to drop in both cases. At both radii in Figure \ref{fig:fidatae}, there is an overall downward trend in the FIDA data which persists from the large $n=2$ chirp at $0.186\usk\second$ until approximately $0.196\usk\second$, after the largest $n=1$ chirp. This trend is also observed in the midplane NC channels, and is matched by a corresponding upward trend in the diagonally-oriented channel, although in the NC data this trend lasts only until around $0.193\usk\second$. The bursts of $n=2$ activity indicated by the dashed blue lines in Figure \ref{fig:fidatae} are correlated with substantially larger changes in the FIDA signal than in the NC signal. A large drop in the core FIDA signal occurs at the time of each of these chirping modes, whereas the outer midplane channel is only noticeably affected by the second burst. This again indicates that the occurrence of each burst of TAE activity is independent of the fast-ion population in at least some of the regions of real and phase space affected by the mode. The relative changes in FIDA signal are much larger than those in the neutron camera signal; this is likely due to the spatial localisation of the FIDA measurements increasing their sensitivity to localised changes in fast-ion density, whereas the neutron camera signal is derived from a more spatially extended region. Note that in none of the FIDA or NC traces do these chirping modes consistently cause either a drop or an increase in signal, even though each of the TAE bursts may individually cause a significant perturbation to the local fast-ion density. This supports the conclusion that in this case, each chirping TAE causes selective redistribution of energetic ions within the plasma rather than coherent outward transport from core to edge as observed in the case of fishbones and, as will be shown later, sawteeth. This observation is consistent with the broader radial structure and coupled poloidal harmonics characteristic of the TAEs, predicted by theoretical calculations and numerical modelling \cite{Cheng1986, Fu1995}, when compared with the core-localised fishbones. 

The contrast between these TAEs, which appear to cause redistribution but no appreciable losses of fast ions, and the chirping TAEs in shot \#29994, which are seen in Figure \ref{fig:adafb} to cause fast-ion losses, may at first seem surprising. The difference is ascribed to the fast-ion slowing down times prevalent in the two cases; the beams start much later in shot \#29902 than in \#29994, and consequently the density is higher and the slowing down time shorter. This weakens the resonant energetic particle drive for the modes, resulting in a slower mode evolution and a weaker effect of the Alfv\'en waves on the fast ions. By contrast, the chirping modes which cause spikes in the passive FIDA light in shot \#29994 evolve extremely rapidly, with the amplitude of the Mirnov coil signal varying from the noise level to or from maximum amplitude within 8-10 wave periods. In shot \#29902, even the most rapidly evolving TAEs take several tens of wave periods to grow and decay.

The spectrograms in Figure \ref{fig:taechirp} show a weak $n=1$ tearing mode at low frequency appearing at around $0.185\usk\second$ and ending abruptly at $0.192\usk\second$. The disappearance of this mode is associated with an internal reconnection event as the on-axis safety factor $q_0$ evolves downward, passing through a value of 2. The core electron temperature measured with the Thomson scattering system also crashes at this time, which would be expected to cause a drop in the core neutron rate on slowing-down timescales ($\sim10\usk\milli\second$); the extremely rapid change in core neutron emissivity, which takes place over much less than a fast-ion slowing down time, along with the fact that the core NC signal subsequently recovers despite the electron temperature remaining suppressed, supports the conclusion that the observed changes are associated with the chirping TAE rather than changes in bulk plasma profiles. The low amplitude of the tearing mode itself, as well as the fact that the nett downward trends in the NC and FIDA signals persist after this mode has been eliminated, mean that changes in these signals are also unlikely to be due to the tearing mode.

The preceding analysis concerning the effects of chirping modes, both at TAE and internal kink frequencies, reveals the potential for joint analysis of FIDA and NC data to constrain models of the fast-ion redistribution resulting from these modes, and to qualitatively diagnose the effects of these modes on the fast-ion distribution. We now summarise the results of a study of the effects of quasi-continuous, or \emph{saturated} MHD modes, in particular the long-lived internal kink mode.

\section{Saturated modes}\label{sec:LLM}
Figure \ref{fig:spectro} shows fishbones evolving later in the discharge into a saturated mode with nearly constant frequency. This mode, observed in many such MAST discharges, is known as the long-lived mode (LLM). Its structure approaches that of the ideal internal kink mode as the minimum of the safety factor approaches unity \cite{Chapman2010}. The core-localised nature of the internal kink mode implies that the effects on the fast-ion population should be similarly localised.

\begin{figure}[h]
\centering
\includegraphics[width=0.5\textwidth]{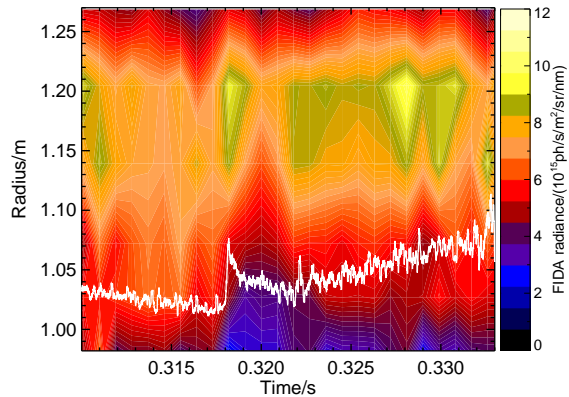}
\caption{FIDA signal ($\lambda=661.0\usk\nano\metre$) from MAST shot \#29914 as a function of radial position and time. Overplotted in white is the amplitude in arbitrary units of the $n=1$ component of the poloidal magnetic field perturbation, signifying the presence of the long-lived mode; growth of the mode, from around $0.318\usk\second$, is observed to cause fast-ion redistribution.}\label{fig:FIDALLM}
\end{figure}
 
\begin{figure*}[h]
\centering
\includegraphics[width=0.8\textwidth]{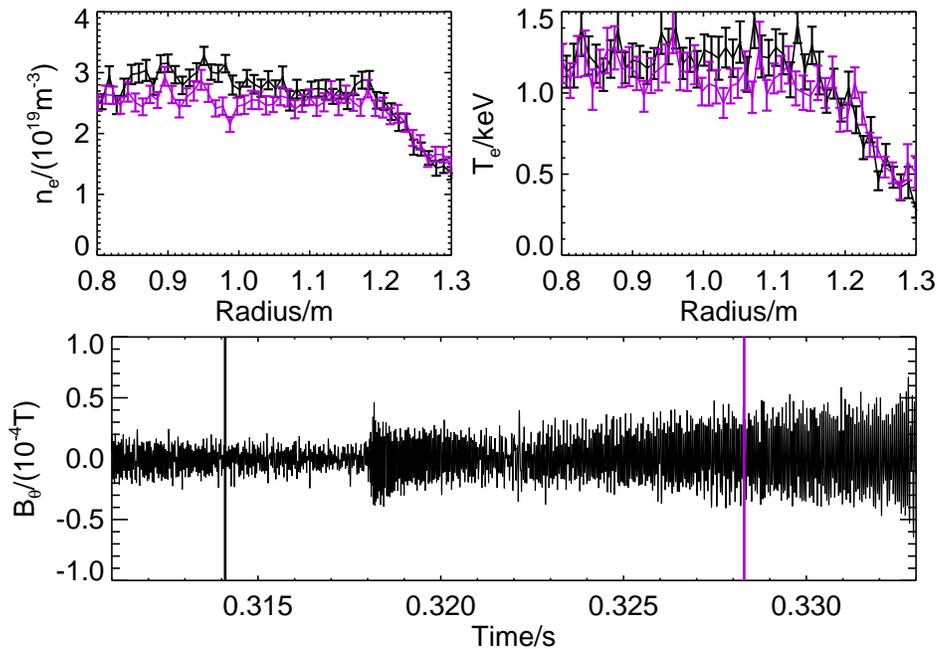}
\caption{Thomson scattering-derived profiles of electron density and temperature from MAST shot \#29914 at two different times, identified by solid vertical lines in the third panel. The third panel shows the $n=1$ component of the perturbed poloidal magnetic field derived from an array of Mirnov coils located on the outboard midplane, allowing the onset and growth of the LLM to be identified.}\label{fig:TSplot}
\end{figure*}

In Figure \ref{fig:FIDALLM}, it is seen that as the LLM grows there is indeed a noticeable suppression of the fast-ion density in the core of the plasma at $R\lesssim1.05\usk\metre$. Redistribution occurs, with the fast ions pushed out to between $1.10\usk\metre$ and $1.25\usk\metre$. Beyond this radius, the signal appears to be largely unaffected. Thomson scattering profiles during the same period reveal very small changes in the core plasma density due to the mode; the electron density in Figure \ref{fig:TSplot} is slightly reduced in the range $0.84\usk\metre<R<1.07\usk\metre$, and the line-integrated density as measured with a laser interferometer drops by approximately 4\% between $0.315\usk\second$ and $0.325\usk\second$. Electron temperature appears to be affected slightly further out, being slightly reduced after mode onset in the range $0.99\usk\metre<R<1.17\usk\metre$. There is no apparent `pile-up' of density outside the affected region, suggesting that the thermal particles are efficiently transported away from the core. Changes in beam deposition cannot therefore account for the observed changes in FIDA signal. Taken together, these results imply that the mode causes fast ions to be redistributed from the core toward the edge of the plasma.

\begin{figure}[h]
	\begin{subfigure}[t]{0.5\textwidth}
	\includegraphics[width=\textwidth]{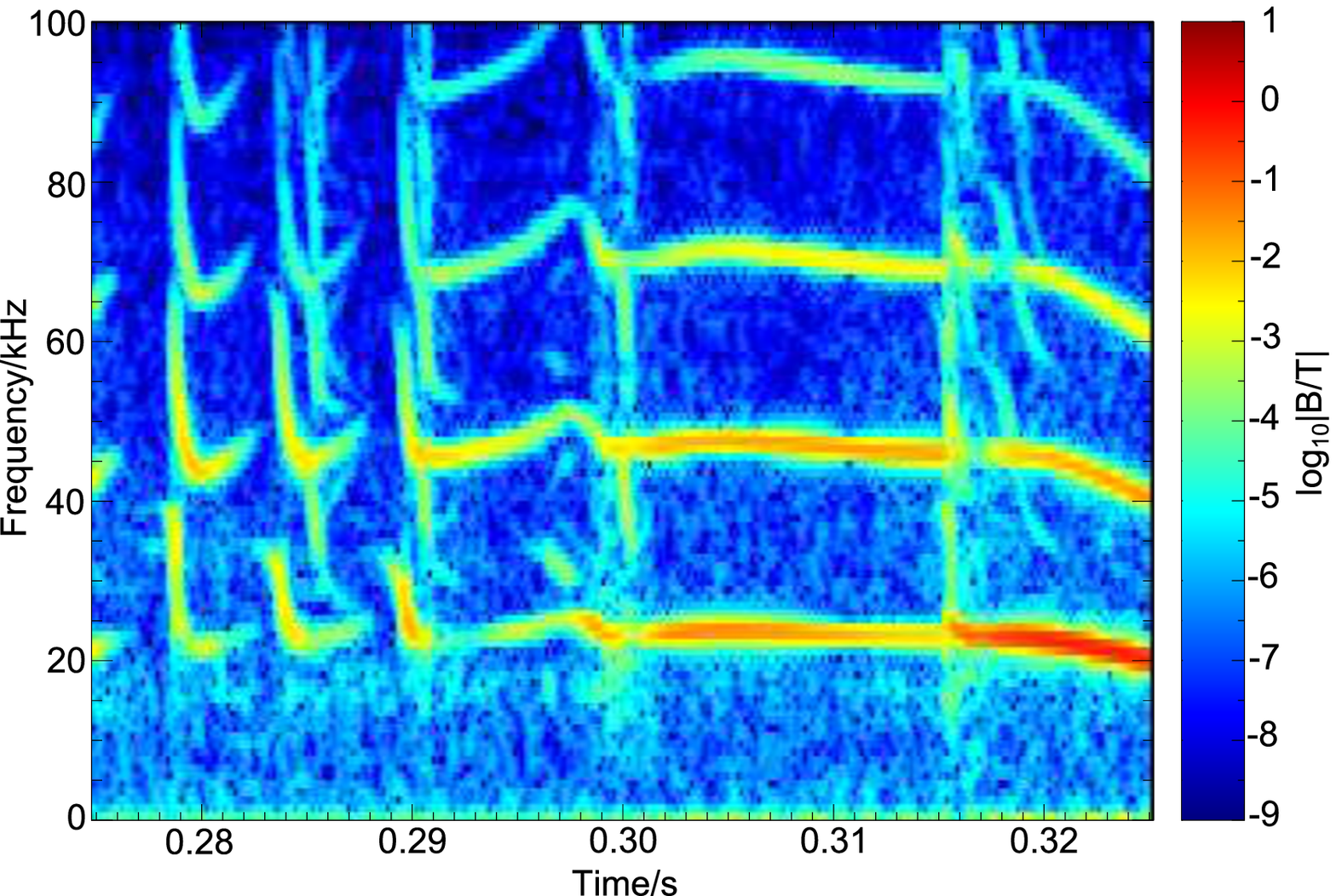}
	\caption{}
	\label{fig:llmspectro}
	\end{subfigure}
	\begin{subfigure}[t]{0.5\textwidth}
	\includegraphics[width=\textwidth]{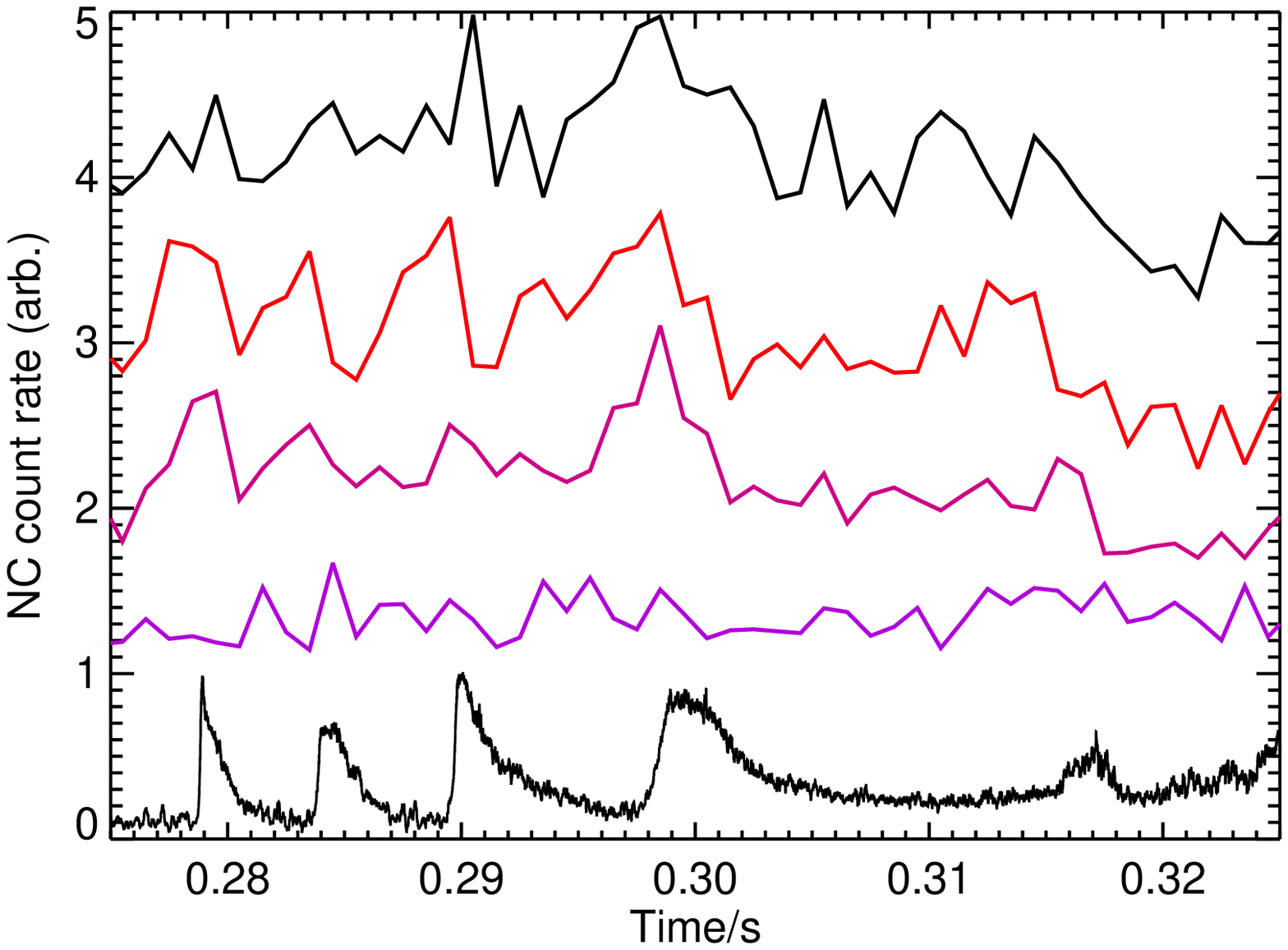}
	\caption{}
	\label{fig:ncllm}
	\end{subfigure}
\caption{(a) Spectrogram of poloidal magnetic field perturbations during MAST shot \#29924 from a Mirnov coil located on the outboard midplane. Three fishbones are clearly identified, followed by a final chirping mode which undergoes a transition to a long-lived mode at approximately $0.30\usk\second$. (b) Time traces of NC signal from the midplane channels for two discharges, \#29924 and \#29928. The time bases of the signals were aligned according to the magnetic activity; the amplitude in arbitrary units of the $n=1$ signal from shot \#29924 is shown for reference. Tangency radii (and SF) are, from top to bottom, $0.88\usk\metre$ (1.0), $1.01\usk\metre$ (0.85), $1.08\usk\metre$ (0.7) and $1.20\usk\metre$ (0.85).}\label{fig:LLM}
\end{figure}

Examining the neutron camera signal from another shot for signs of redistribution due to the LLM, the case is similarly clear-cut. The signal from the three innermost channels in Figure \ref{fig:ncllm} is observed to be strongly depleted from the onset of the mode at around $0.30\usk\second$, although in this case the outer channel with a tangency radius of $R=1.20\usk\metre$ does not detect a corresponding increase in signal. The amplitude of the LLM is very similar in the shots considered in Figures \ref{fig:FIDALLM} and \ref{fig:LLM}, but the FIDA diagnostic detects an increase in signal in shot \#29924 further out than was observed in shot \#29914; the FIDA signal at $R=1.20\usk\metre$ barely shows any nett change during the period examined in Figure \ref{fig:LLM}, but there is a noticeable increase in signal at $R\geq1.27\usk\metre$ once the LLM appears. It is not entirely clear why the region to which fast ions are redistributed by the mode should be different when considering these two very similar discharges, but it is noted once again that an observation with one fast-ion diagnostic is supported by the other. Combining observations from the neutron camera and FIDA spectrometer places strong constraints on the behaviour of fast ions affected by MHD modes.

Previous observations of the effects of the LLM in MAST have independently considered the effects of this mode on the global neutron rate and the NC and FIDA signals \cite{Michael2013c, Cecconello2012}. Incorporating the analysis of data from several diagnostics in the present work once again demonstrates the power of this approach in developing a consistent picture of the effects of MHD modes on fast ions. The previous study of the effect of the LLM on NC profiles established a strong suppression of emission from the core, which was attributed to radial redistribution after consideration of changes in bulk plasma profiles \cite{Cecconello2012}. Adding FIDA data to the analysis, as has been done in the present study, firmly establishes this model as well as providing a measurement of enhanced signal in the radial region to which the fast ions are redistributed. Ongoing modelling work seeks to determine the form of the helical equilibrium associated with the presence of the long-lived mode and the consequences for the fast-ion distribution \cite{Pfefferle}; the development of synthetic diagnostics for the NC and FIDA systems will allow experimental constraints to aid this modelling effort. 

\section{Sawteeth and edge-localised modes}
The final two instabilities considered in this study are, unlike chirping modes and the LLM, encountered in MAST plasmas even in the absence of neutral beam injection. The observations reported here are notable in that they reveal an effect on the fast-ion population both from the individual occurrence of each of these modes, and from events in which the modes are coupled, or synchronised. Edge-localised modes (ELMs) have been observed to correlate with sawteeth in TCV under certain conditions \cite{Martin2002}; this behaviour has also been observed in MAST but has not been systematically studied.

\begin{figure}[h]
\centering
\includegraphics[width=0.95\textwidth]{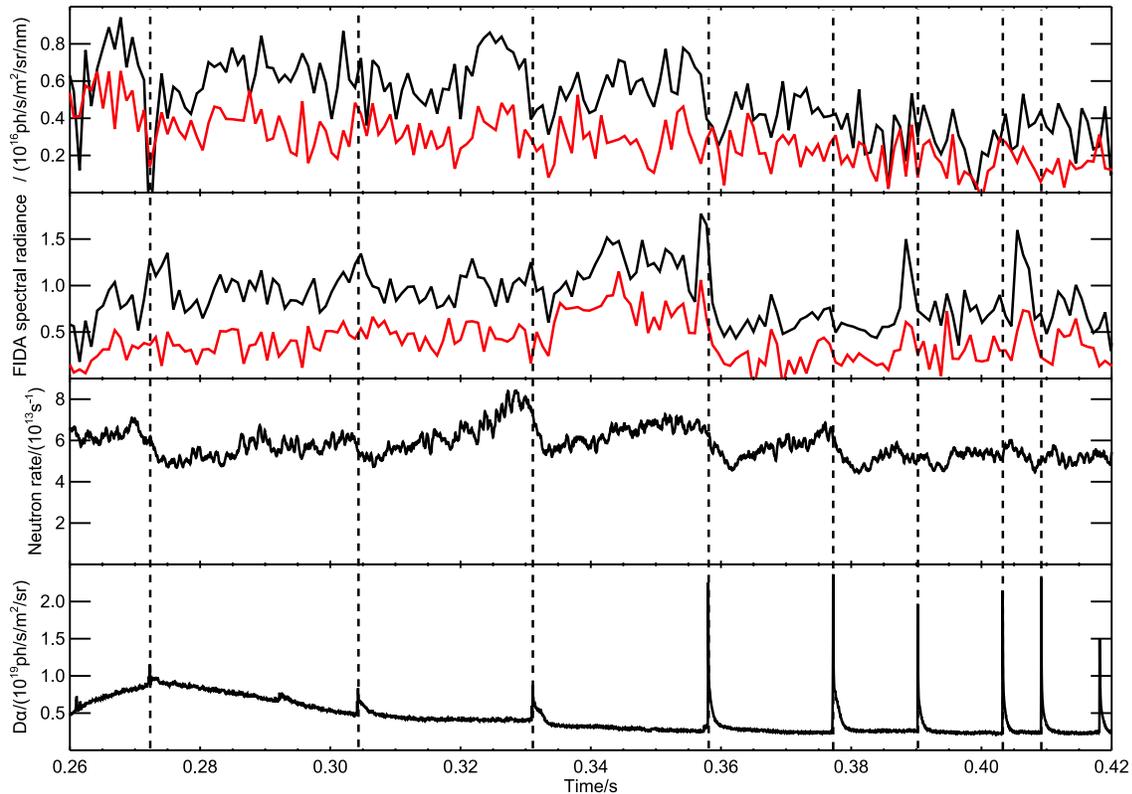}
\caption{FIDA data from a period during MAST shot \#30006 in which sawteeth and ELMs were observed. Top panel: $R=1.03\usk\metre$; $\lambda=660.7\usk\nano\metre$ (black) and $661.4\usk\nano\metre$ (red). Second panel: $R=1.27\usk\metre$; same wavelengths. Third panel: global neutron rate observed by the fission chamber. Bottom panel: D$\alpha$ signal observed by a chord viewing bulk plasma volume from outboard midplane. The two wavelengths chosen for the FIDA time traces correspond to minimum fast-ion energies of $46.1\usk\kilo\electronvolt$ (black) and $61.2\usk\kilo\electronvolt$ (red). Dashed vertical lines mark the onset of sawteeth and ELMs as observed in the edge D$\alpha$ trace.}
\label{fig:FIDA_ELM}
\end{figure}

MAST discharge \#30006 underwent a transition to H-mode during a period in which sawteeth were active. The L-H transition itself was associated with a large sawtooth, and the first two ELMs also exhibited core electron temperature crashes characteristic of sawteeth, as well as the typical sawtooth precursor (a rapidly-growing $n=1$ kink mode) appearing in the magnetic pickup coil trace $1-2\usk\milli\second$ before the ELM. Figure \ref{fig:FIDA_ELM} shows the FIDA and fission chamber signal evolution during this period; three sawteeth up to and including the L-H transition at $0.332\usk\second$ are apparent as small perturbations to the edge D$\alpha$ trace in the bottom panel and significant drops in the fission chamber signal in the third panel. These are followed by a series of ELMs, visible as large spikes in the D$\alpha$ trace; the first two of these in particular are correlated with very large drops in the global neutron rate. These two ELMs are each associated with a simultaneous sawtooth crash; the reduced temperature in the plasma core causes the slowing-down time of the beam ions to decrease, reducing the total neutron yield. The fact that changes in the global neutron rate occur over timescales somewhat less than the slowing down time suggests that redistribution of fast ions may contribute to these changes in neutron rate. From inspection of the fission chamber data alone, and without further detailed modelling, these data are however insufficient to confirm whether redistribution is taking place.

Examination of the FIDA data in Figure \ref{fig:FIDA_ELM} reveals a broadly consistent picture of the effects of ELMs and sawteeth on the fast-ion density at the selected radii and wavelengths, with a couple of notable exceptions. The sawteeth, including those associated with the first two ELMs, cause noticeable drops in the signal in the mid-energy core channel (black trace, top panel). Transient increases in signal are simultaneously observed in the mid-energy edge channel (black trace, second panel). General points to note are that mid-energy signals (black traces) are more strongly affected than high-energy signals (red traces), and that the signal nearer the edge (second panel) is usually more strongly affected by ELMs than the signal in the core (top panel) whereas the opposite applies to sawteeth. The high-energy core signal shows significant drops associated with the first and third saweeth. Transient spikes are seen in the mid-energy edge signal immediately before the first, third and fifth ELMs, but the occurrence of these spikes appears to have no bearing on the nett evolution of the signal around the time of the event; the signal drops strongly around the time of the first ELM, but is on average much more weakly perturbed by subsequent ELMs despite the transient increase. One particular point to note is that the effect of the second ELM, or rather its associated sawtooth, on the FIDA signal in all channels is much weaker than that of the first ELM, despite the fact that these events have a similar impact on the global neutron rate. The reason for this apparent discrepancy is unclear.

\begin{figure}[h]
\centering
\includegraphics[width=0.5\textwidth]{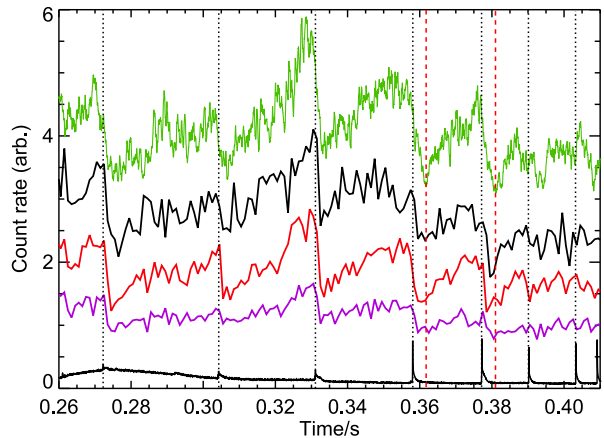}
\caption{Fission chamber (green) and neutron camera (black, SF = 0.9; red, SF = 0.55; purple, SF = 0.35) signals during MAST discharge \#30006. The tangency radii of the inner and outer chords are $R=0.84\usk\metre$ and $R=1.04\usk\metre$. Black and red traces are from the inner and outer midplane chords, while the purple trace is from the inner, diagonally-oriented chord. ELMs are identified as spikes in the trace of edge D$\alpha$ light, overplotted in black. Dotted vertical lines mark the onset of sawteeth and ELMs as seen in the edge D$\alpha$ trace, while the red, dashed vertical lines demarcate the end of the drops in fission chamber count rate associated with the first two ELMs.}
\label{fig:NC_ELM}
\end{figure}

Neutron camera data are available from the same shot, and exhibit drops in count rate associated with each of the sawteeth, including those with ELMs. It is interesting to note that the drop in NC count rate is slightly faster in many cases than the drop in fission chamber count rate for the sawteeth which are correlated with ELMs; the red, dashed vertical lines in Figure \ref{fig:NC_ELM} indicate the end of the drop in fission chamber signal, but in most cases the neutron camera signal is already increasing by the time the fission chamber signal reaches a minimum. This is consistent with the understanding that the sawtooth rapidly redistributes fast ions to the edge of the plasma, where they make a smaller contribution to the NC signal but continue to contribute to the global neutron rate until they have slowed down. The competing processes of slowing down of redistributed fast ions, which reduces the neutron rate, and deposition of high energy ions in the plasma core, which increases the neutron rate, govern the post-crash behaviour of the neutron emission. The fact that this difference in timescales is only observed in cases where the sawtooth accompanies an ELM may be a result of the higher density in the outer part of the plasma in H-mode; the rate of beam-target fusion reactions is proportional to both the beam-ion density and thermal target density, so the rate will not drop as quickly if the fast ions are redistributed to a region where the thermal ion density is high. Note also that in this case, the core midplane channel which most closely matches the tangency radius of the core FIDA channel (red trace) \emph{does} exhibit a significant drop associated with the second ELM; this suggests that the FIDA measurements, which do not show such a change, may be missing part of the fast-ion distribution in real or velocity space which is strongly affected by this sawtooth.
\begin{figure}[h]
	\begin{subfigure}[t]{0.5\textwidth}
	\centering
	\includegraphics[width=\textwidth]{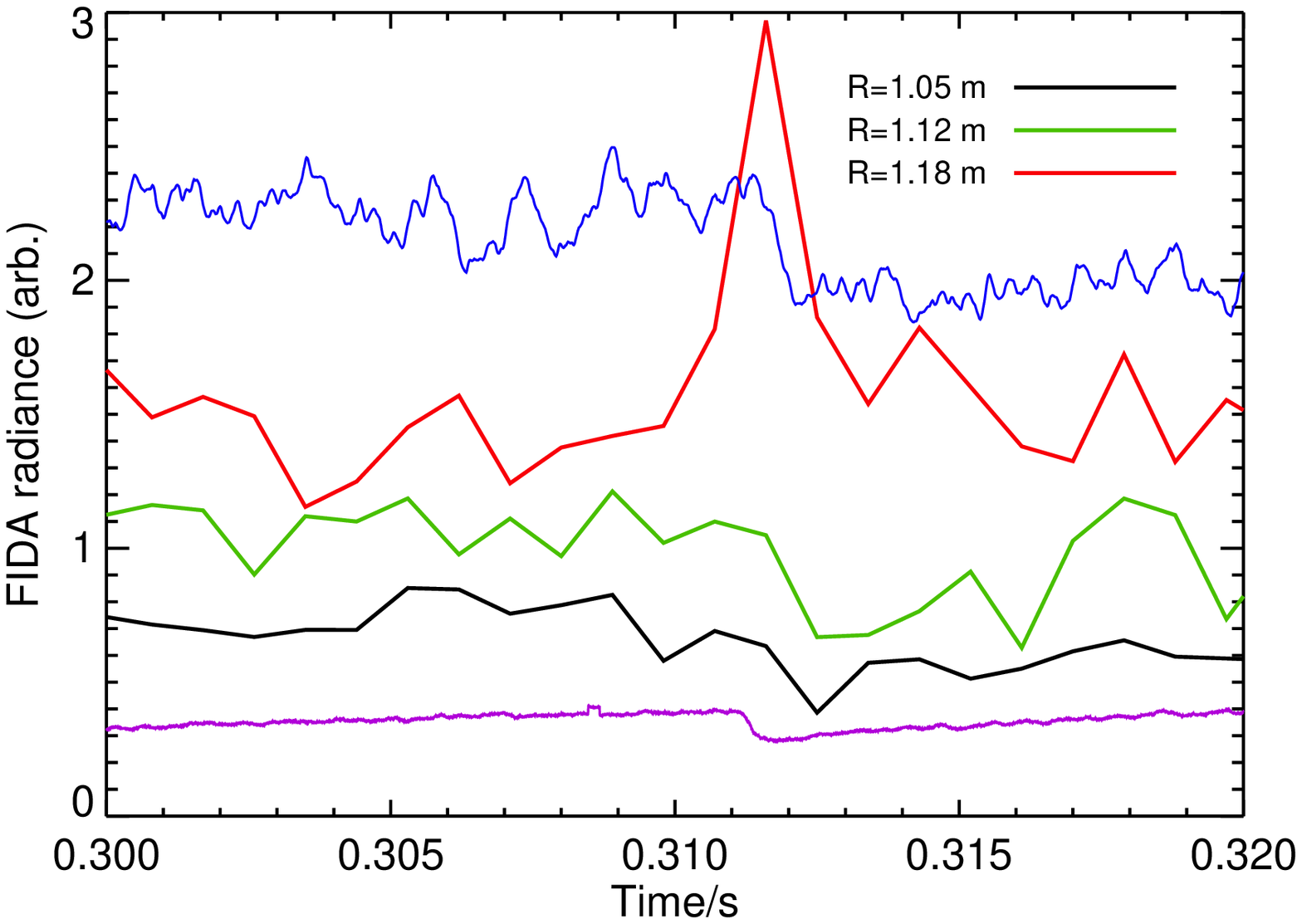}
	\caption{}
	\label{fig:FIDA_ST}
	\end{subfigure}
	\begin{subfigure}[t]{0.5\textwidth}
	\centering
	\includegraphics[width=\textwidth]{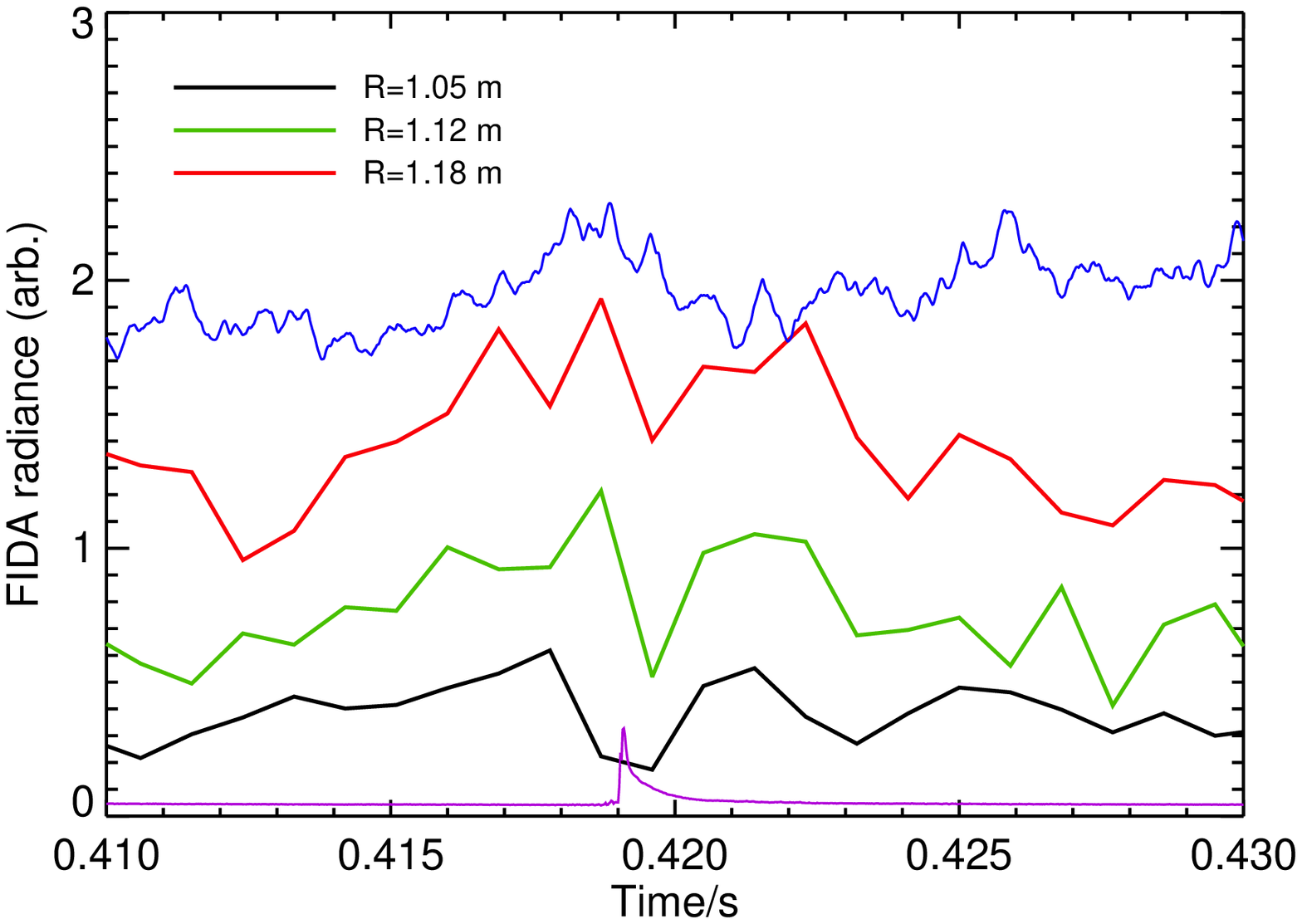}
	\caption{}
	\label{fig:FIDA_ELM2}
	\end{subfigure}
\caption{Evolution of FIDA signal at three different radii, at a wavelength of $660.8\usk\nano\metre$, during MAST shot \#30147. (a) shows the signal around the time of a sawtooth crash while the discharge is in L-mode. (b) shows the signal from the same channels around the time of an ELM with very weak signs of an associated sawtooth crash. The fission chamber signal, scaled to fit on the figure, is overplotted in blue while the purple traces in (a) and (b) show the intensity of soft X-ray and edge D$\alpha$ light respectively, allowing the sawtooth and ELM to be identified. Scaling factors in the two figures are: 0.88 (black); 1.12 (green); 1.23 (red).}
\label{fig:FIDA_comparison}
\end{figure}

Finally, in Figure \ref{fig:FIDA_comparison}, we make a side-by-side comparison of the evolution of FIDA and fission chamber signals during a sawtooth versus an ELM (these data are from a shot with the same plasma current and beam power as \#30006 but with much higher density). In the case of the sawtooth in Figure \ref{fig:FIDA_ST}, drops in signal are observed in the inner two channels while the outer channel shows a transient spike in FIDA radiance at the time of the crash. The inversion radius, determined by the transition from `drop' to `spike' dynamics in the FIDA signal, therefore lies between $1.12\usk\metre$ and $1.18\usk\metre$ on the machine midplane. This is consistent with the sawtooth inversion radius determined from the soft X-ray emission, which is found to be between $1.14\usk\metre$ and $1.22\usk\metre$; combining these constraints puts the inversion radius between $1.14\usk\metre$ and $1.18\usk\metre$. The behaviour observed in the FIDA data in this case is consistent with that from the neutron camera as seen in Figure \ref{fig:NC_ELM}, and with previous findings on other tokamaks \cite{Muscatello2012}; sawteeth redistribute passing fast ions from inside to outside the inversion radius. Note also that these sawteeth are observed to cause large spikes in the passive FIDA signal in channels viewing the edge of the plasma, in a similar manner to the coupled TAE-fishbones, but in this case the spikes are observed right up to the beam injection energy. No such spikes are observed above the injection energy, which would be the case if the increase in signal were caused by a sudden increase in bremsstrahlung emission due to temperature or density perturbations. This suggests that sawteeth cause even the most energetic ions to be lost from the plasma, which is consistent with their large effect on the global neutron rate. 

The ELM in Figure \ref{fig:FIDA_ELM2} by comparison does not exhibit any inversion radius; the FIDA signal is perturbed, undergoing a transient drop in each channel, but the recovery is quicker than in the case of the sawtooth and the signal in each channel appears to be affected in much the same way. Unlike in the case of the sawtooth, where the fission chamber signal undergoes a rapid nett drop at the time of the crash, the perturbation to the fission chamber signal around the time of the ELM consists of a rise in signal before the event, followed by a relatively slow decrease, then a recovery in the subsequent $5\usk\milli\second$ period. These FIDA and fission chamber observations imply that rather than a large-scale redistribution of fast ions from inside to outside the inversion radius as observed in the case of the sawtooth, the components of the field associated with the ELM which penetrate deep into the plasma serve only to redistribute fast ions relatively weakly and in a rather more localised sense. Note also that none of the FIDA or NC channels, or the fission chamber signal, provide evidence that the effect of coupled sawteeth and ELMs on fast-ion confinement is any stronger than the effect of sawteeth alone. One final point to note is that while sawteeth, being internal reconnection events, cause the positions of flux surfaces within the plasma to shift rapidly, the large changes in global neutron rate observed at each event suggest that the changes in FIDA and NC signal are likely to be due predominantly to redistribution and loss of fast ions rather than the resultant changes in the diagnostics' viewing geometry with respect to the magnetic equilibrium.

The extensive fast-ion diagnostic capability on MAST provides the possibility to constrain modelled fast-ion distributions. The Monte Carlo beam-ion module NUBEAM \cite{Pankin2004}, incorporated into the global, time-dependent transport code TRANSP, models the deposition of injected beam neutrals and tracks the fast ions until they either thermalise or are lost from the plasma. NUBEAM outputs the fast-ion distribution as a function of $R$, $Z$, energy $E$ and pitch $p=v_\parallel/v$ at selected times. The comparison between measured and modelled FIDA and neutron emission is presented in the following section.

\section{Modelling the fast-ion distribution}\label{sec:model}
\begin{figure}[h]
\centering
\includegraphics[width=0.50\textwidth]{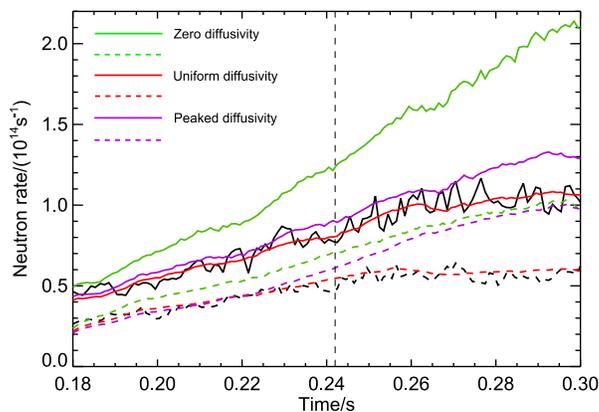}
\caption{Comparison of total neutron rates from low-density 1- and 2-beam MAST discharges \#29222 (dashed lines) and \#29221 (solid lines). Also shown are the predicted neutron rates from TRANSP under assumptions of zero anomalous fast-ion diffusion (green), spatially-uniform anomalous diffusion (red) and centrally-peaked anomalous diffusion inside $\psi_\mathrm{N}=0.1$ (purple). The dashed vertical line shows the time selected for radial profiles in Figures \ref{fig:29222FIDA} and \ref{fig:29221FIDA}. In the uniform diffusivity case, anomalous diffusivity values at the selected time are (a) $1.25\usk\metre^2\second^{-1}$ and (b) $2.25\usk\metre^2\second^{-1}$. In the peaked diffusivity case, central values are (a) $2.86\usk\metre^2\second^{-1}$ and (b) $5.14\usk\metre^2\second^{-1}$.}\label{fig:TRneut}
\end{figure}

In a recent experiment, a given $800\usk\kilo\ampere$ double-null-diverted scenario was run twice, with 1-beam heating in one shot and 2-beam heating in the other. At the relatively low density chosen, these shots exhibited strong MHD activity in the form of chirping and long-lived modes at both levels of beam power. The resulting fast-ion redistribution is seen clearly in Figure \ref{fig:TRneut}; both the 1-beam and 2-beam discharges exhibit a neutron rate which is significantly suppressed compared to the expected rate modelled using TRANSP (green traces). Introducing anomalous fast-ion diffusivity into the TRANSP simulation is necessary to recover the observed neutron rate; the red traces in Figure \ref{fig:TRneut} show the neutron rate obtained by applying a time-dependent, spatially-uniform diffusivity, whereas the purple traces show the modelled neutron rate when diffusivity is applied with the same time dependence and ratio between the values in the 1-beam and 2-beam shots as applied in the uniform diffusivity case, but with a profile which is strongly peaked in the core (diffusion coefficients $D(\psi_\mathrm{N}=1)=0.25D(\psi_\mathrm{N}=0)=0.25D(\psi_\mathrm{N}=0.1)$, with the diffusivity falling off linearly between $\psi_\mathrm{N}=0.1$ and $\psi_\mathrm{N}=1$). This model was chosen due to the fact that the experimental observations relevant to TAEs, which occurred in these discharges, indicate that some redistribution occurs even in the outer part of the plasma. Setting the diffusivity to zero outside a certain radius as has been done in previous studies \cite{Michael2013c} was therefore not expected to represent the true situation accurately. This choice is somewhat arbitrary however, since transport of fast ions by MHD modes is not believed to be truly diffusive in nature; anomalous diffusion simply represents a convenient \emph{ad hoc} model which may be applied in the absence of a first-principles model of transport.

\begin{figure}[h]
	\begin{subfigure}[t]{0.5\textwidth}
	\centering
	\includegraphics[width=\textwidth]{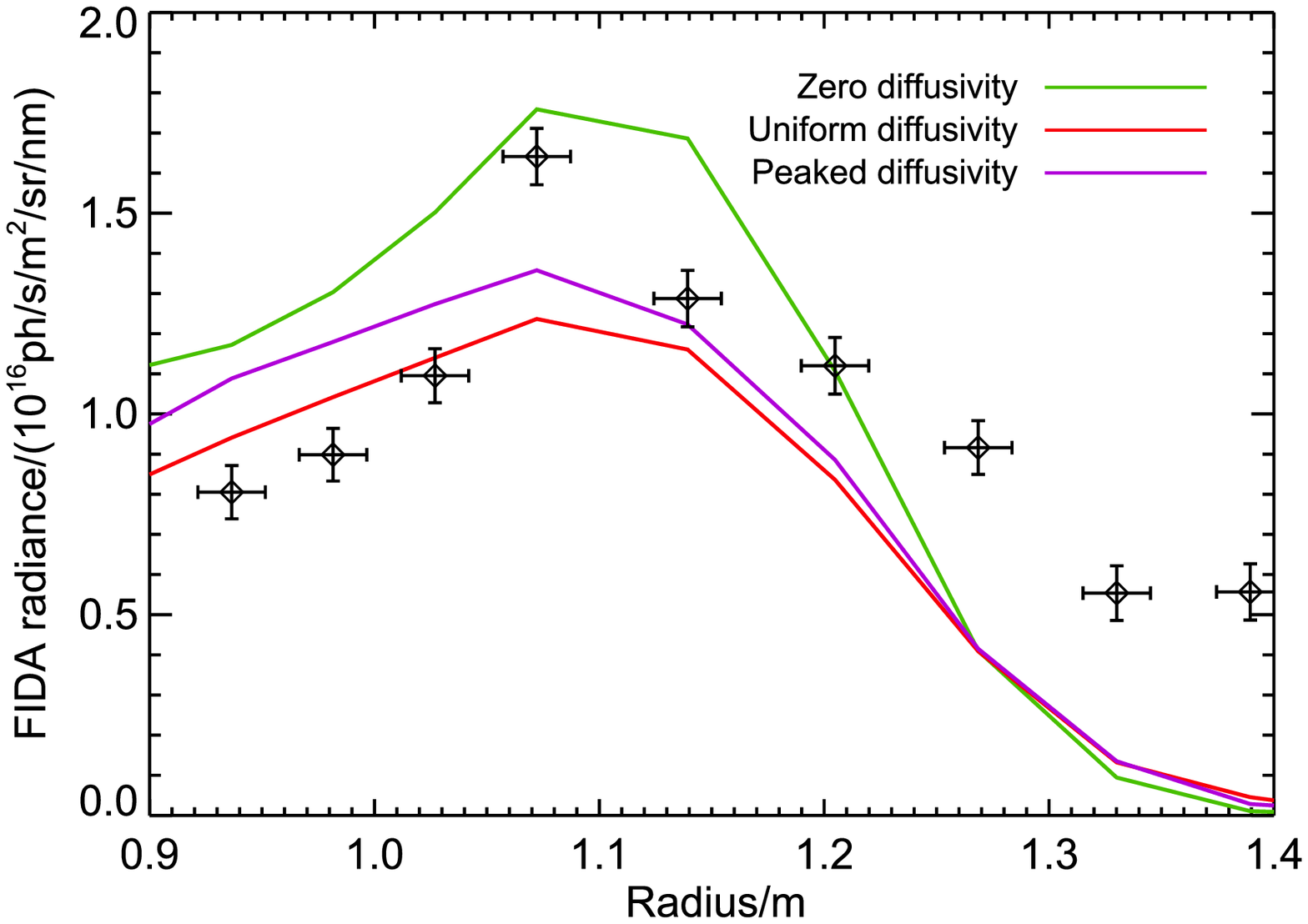}
	\caption{}
	\label{fig:29222FIDA}
	\end{subfigure}
	\begin{subfigure}[t]{0.5\textwidth}
	\centering
	\includegraphics[width=\textwidth]{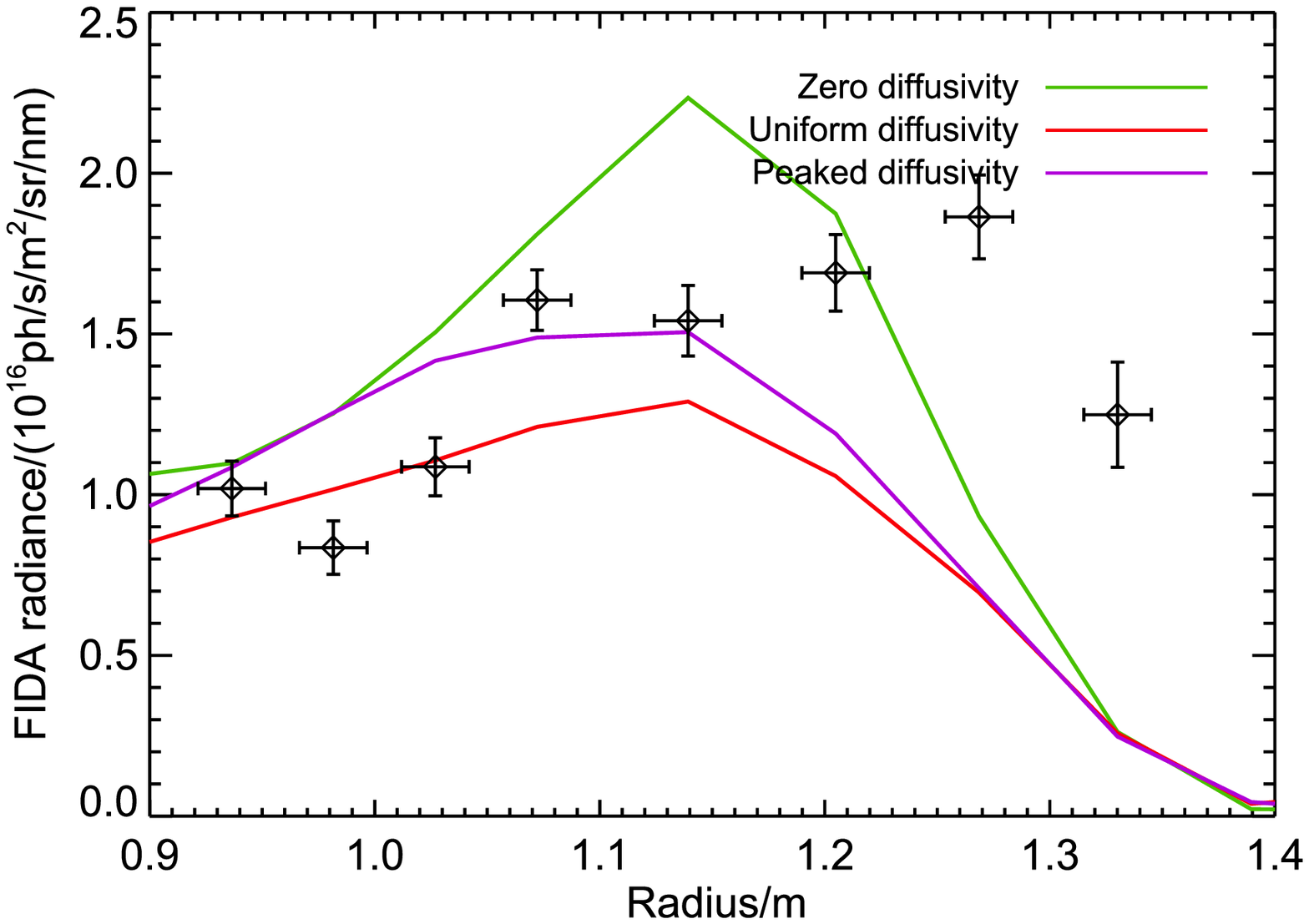}
	\caption{}
	\label{fig:29221FIDA}
	\end{subfigure}
\caption{Radial profiles of FIDA signal at $t=0.242\usk\second$, $\lambda=660.6\usk\nano\metre$ from MAST discharges \#29222 (a) and \#29221 (b). Overlaid are the signals modelled using the FIDAsim code based on the TRANSP fast-ion distributions assuming three different levels of anomalous fast-ion diffusion (the three levels being the same as those shown in Figure \ref{fig:TRneut}).}
\label{fig:2122FIDA}
\end{figure}

The fast-ion distribution generated by the NUBEAM module within TRANSP for each of these runs was output at $t=0.242\usk\second$, indicated by the dashed vertical line in Figure \ref{fig:TRneut}. This output was then processed using the Monte Carlo neutral deposition code FIDAsim \cite{Heidbrink2011}, which tracks injected neutrals and fast ions which have undergone re-neutralisation and models the collisional-radiative processes affecting each of these. The code outputs FIDA spectra at each radial position at which lines of sight intersect the neutral beam. Comparisons between the measured and modelled FIDA profiles are shown in Figure \ref{fig:2122FIDA}. It is seen that in neither case does the model of anomalous diffusion which best fits the measured neutron rate, namely that of spatially uniform diffusivity, give a particularly good match to the measured FIDA signal. There is a tendency for all purely diffusive models of transport to overestimate the FIDA signal in the core of the plasma and underestimate the signal further out. Note that the signal in Figure \ref{fig:29221FIDA} is affected by a fishbone which occurs at approximately $0.241\usk\second$; the profile before the fishbone onset is more peaked, and is matched more closely by the model assuming zero anomalous diffusion which is seen in Figure \ref{fig:TRneut} to fail to reproduce the observed neutron rate.

\begin{figure}[h]
\centering
\includegraphics[width=0.5\textwidth]{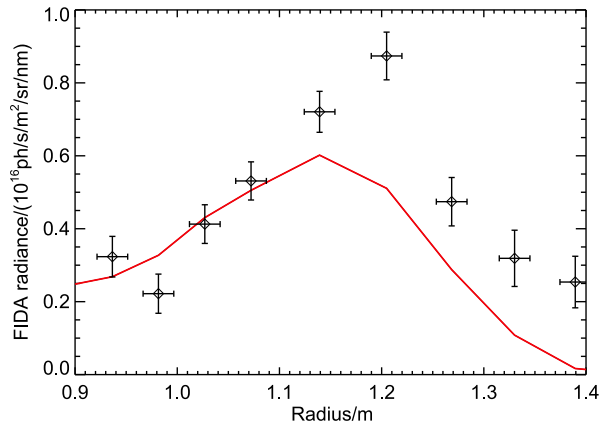}
\caption{Radial profile of FIDA signal at $t=0.195\usk\second$, $\lambda=660.6\usk\nano\metre$ from MAST discharge \#29904. Overlaid in red is the signal modelled using the FIDAsim code based on the TRANSP fast-ion distribution assuming zero anomalous diffusion.}\label{fig:29904FIDA}
\end{figure}

Confidence in the FIDA results is reinforced by comparing modelled results with experimental data from an MHD-quiescent discharge. MAST shot \#29904 featured $1.5\usk\mega\watt$ of NBI power injected into a high-density plasma which was otherwise similar to shots \#29221 and \#29222. Low beam power and high density weaken the drive for energetic particle instabilities. As a result, the plasma exhibited periods with no appreciable MHD activity. The neutron rate and stored energy predicted by TRANSP are close to the measured neutron rate and the stored energy reconstructed by EFIT. Agreement between the TRANSP-predicted and experimentally measured neutron rate is better than 20\% throughout the period of the discharge which was analysed; this agreement improves to 5\% for the TRANSP versus EFIT estimates of stored energy. An example of a FIDA profile from this shot is shown in Figure \ref{fig:29904FIDA}. At this time during the shot, the TRANSP-predicted and experimentally measured neutron rates agree to within 5\%.

There is a data point at $R=1.21\usk\metre$ which departs strongly from the general shape of the profile; this arises due to the fact that at this radius, the selected wavelength coincides with the Doppler shifted beam emission of the small amount of hydrogen in the neutral beam. The spectral contamination also affects data points either side of this radius to a lesser degree, increasing the measured signal at $R=1.14\usk\metre$ and $R=1.27\usk\metre$ above the true active FIDA signal. Setting aside this undesirable contamination of the signal, the general agreement between the measured and modelled FIDA profile is reasonably good with regard to both shape and absolute magnitude in the core of the plasma. This agreement is weaker toward the outer edge of the plasma, where the signal is underestimated by the modelling codes, although the shape of the profile is matched reasonably well in this case. This consistent discrepancy in both MHD-active and MHD-quiescent plasmas hints either at a systematic error in background subtraction of the FIDA signal at these radii, or at some physical process which is not captured by the codes. These may include redistribution of fast ions by toroidal field ripple or residual error fields in the outer part of the plasma. Investigation into the cause of the systematic discrepancy is ongoing. Previous investigation of the radial profile of neutron emission with the NC has not uncovered any substantial inconsistency between measurements and modelling in the case of MHD-quiescent discharges \cite{Turnyanskiy2013}, but the neutron camera signal outside a radius of approximately $1.2\usk\metre$ is negligible; it is only outside this radius that the FIDA diagnostic detects an appreciable discrepancy. In addition, the FIDA diagnostic provides an additional degree of freedom over that available to the neutron camera, in that the analysis may be performed at different wavelengths.

\section{Conclusions \& future work}\label{sec:conc}
Recent experiments on MAST provided the opportunity to field several fast-ion diagnostics simultaneously. The effects of frequency-chirping MHD modes, saturated internal kink modes, sawteeth and ELMs on the FIDA signal and neutron emission from the plasma have been studied in detail. Joint analysis of data from the FIDA spectrometer and neutron camera has yielded the following observations:
\begin{itemize}\itemsep0pt
\item Core-localised drops in the signal due to fishbones indicate that fast ions are redistributed from the centre of the plasma, with losses occurring when the fishbones have a `TAE-like' nature. There is some evidence that the relative change in fast-ion density in the core scales with the peak time derivative of the magnetic field perturbation.
\item Chirping TAEs with different toroidal mode numbers appear to affect fast ions at different radii, with $n=1$ TAEs principally affecting the core fast-ion density and $n=2$ TAEs affecting fast ions further out. Numerous, closely-spaced chirping TAEs can cause nett suppression of the fast-ion density on the midplane, but overall loss of confinement is not observed. More strongly driven TAEs with large amplitude and rapid temporal evolution \emph{are} however observed to cause small levels of fast-ion losses.
\item Radial redistribution is associated with the long-lived mode, whereby fast ions are removed from the core and pushed out to larger radii. The region from which fast ions are expelled coincides with that from which a small amount of the bulk plasma is removed.
\item Sawteeth cause rapid drops in the neutron camera and FIDA signals in the core, and an increase in FIDA signal outside the sawtooth inversion radius. The inversion radius is determined self-consistently from the FIDA signal and the soft X-ray emission. The large drops in global neutron rate, and long recovery time, suggest that fast ions are lost from the plasma due to the sawteeth.
\item ELMs associated with sawteeth are correlated with rapid drops in the core neutron camera signal and a marginally slower decrease in the volume-integrated fission chamber signal, suggesting redistribution from core to edge due to the sawtooth. These events deplete the edge FIDA signal from fast ions at all energies, but do not affect the most energetic ions in the core of the plasma.
\item ELMs with no associated sawtooth cause small perturbations to the FIDA signal at all radii, but are not associated with a significant nett depletion. The same applies to the global and localised neutron rates observed by the fission chamber and neutron camera, indicating weak redistribution on small spatial scales.
\end{itemize}

Two key results from this study which should be highlighted due to their potential impact beyond MAST are the following: firstly, fishbones are seen to have a strong deleterious effect on the density of passing fast ions throughout the core of the plasma, and to be driven unstable in a `limit cycle' by these passing particles. This is consistent with an earlier study of the effects of fishbones on fast ions in MAST \cite{Jones2013}, and supports the interpretation of these modes in accordance with the model of Kolesnichenko \etal \cite{Kolesnichenko2006} which was specifically developed with consideration of the prevailing conditions in spherical tokamaks. A recent numerical study by Wang \etal \cite{Wang2013}, aimed at developing an understanding of fishbones in spherical tokamaks, highlighted the similarities of the modes observed in their simulations to the conventional precession-resonant fishbones first observed in PDX in 1983 \cite{McGuire1983, White1983, Hsu1994}. The results obtained in the present study however suggest that, at least under the conditions commonly found in MAST plasmas, resonances with circulating energetic ions may be dominant in determining fishbone stability and consequent fast-ion transport. The second key point raised by this empirical study is that strongly-driven, large-amplitude chirping TAEs do not necessarily cause significant overall loss of confinement even when the ratio of fast-ion Larmor radius to plasma minor radius is large, as is the case in MAST ($\rho/a\sim0.15$). Losses of fast ions are observed in some cases, but the weak effect on the global neutron rate and FIDA signal from the confined fast ions suggests that these losses are small compared to the total fast-ion content of the plasma. This result suggests that TAEs driven unstable by fusion alphas in ITER, the normalized orbit width of which is similar to that of beam ions in MAST, may not necessarily cause significant degradation of confinement. Attention must however be paid to potential localised heat loads on the first wall, as even a small percentage of the fast-alpha energy content in ITER could cause damage if delivered to a limited area within a short space of time.

Attempts to model the redistribution of fast ions in MAST by applying anomalous diffusion within TRANSP have proceeded with some success, but a fully consistent model which simultaneously matches all observable parameters remains elusive. Time-dependent, spatially-uniform anomalous diffusion is able to reproduce the global neutron rate in cases in which redistribution due to MHD modes is observed, but matching the modelled FIDA signal and neutron emissivity profile may require the inclusion of diffusion and advection operators which depend on spatial position, particle energy, pitch angle and time. Since there is only limited theoretical understanding of the appropriate choices for these operators \cite{Pinches2012}, they must necessarily be applied on an \emph{ad hoc} basis when employing codes such as TRANSP. Developing a fully self-consistent basis for the choice of diffusion and advection coefficients will ultimately aid progress towards a first-principles understanding of fast-ion transport due to MHD modes.

\ack The authors are grateful to Eric Fredrickson for discussion and participation in experiments pertaining to TAE avalanches. This work was part-funded by the RCUK Energy Programme under grant EP/I501045 and the European Communities under the contract of association between EURATOM and CCFE. To obtain further information on the data and models underlying this paper, please contact PublicationsManager@ccfe.ac.uk. The views and opinions expressed herein do not necessarily reflect those of the European Commission.

\section*{References}
\bibliographystyle{unsrt}
\bibliography{library}

\end{document}